\documentclass[fleqn,usenatbib]{mnras}

\usepackage{newtxtext,newtxmath}

\usepackage[T1]{fontenc}
\usepackage{ae,aecompl}

\usepackage{graphicx}	
\usepackage{amsmath}	
\usepackage{amssymb}	
\usepackage{float}
\usepackage{rotating}

\usepackage[caption=false]{subfig}

\usepackage{etoolbox}
\makeatletter
\patchcmd\@combinedblfloats{\box\@outputbox}{\unvbox\@outputbox}{}{%
  \errmessage{\noexpand\@combinedblfloats could not be patched}%
}%
\makeatother

\newcommand\nodata{ ~$\cdots$~ }

\title[Swift\,J1753.5-0127 as a Z Cam analogue]{The Curious Case of Swift J1753.5-0127: A Black Hole Low-mass X-ray Binary Analogue to Z Cam Type Dwarf Novae
}

\author[A. W. Shaw et al.]{
A. W. Shaw,$^{1}$\thanks{E-mail: aarran@ualberta.ca (AWS)}
B. E. Tetarenko,$^{1,2}$\thanks{E-mail: btetaren@umich.edu (BET)}
G. Dubus,$^{3}$
T. Din\c{c}er,$^{4}$
J. A. Tomsick$^{5}$
P. Gandhi,$^{6}$
\newauthor{ R. M. Plotkin$^{7}$
and D. M. Russell$^{8}$}
\\
$^{1}$Department of Physics, University of Alberta, CCIS 4-181, Edmonton, AB T6G 2E1, Canada\\
$^{2}$Department of Astronomy, University of Michigan, 1085 South University Avenue, Ann Arbor, MI 48109, USA\\
$^{3}$Univ. Grenoble Alpes, CNRS, Institut de Plan\'{e}tologie et d'Astrophysique de Grenoble (IPAG), 38000 Grenoble, France\\
$^{4}$ Department of Astronomy, Yale University, P.O. Box 208101, New Haven, CT 06520-8101, USA\\
$^{5}$Space Sciences Laboratory, 7 Gauss Way, University of California, Berkeley, CA 94720-7450, USA\\
$^{6}$Department of Physics and Astronomy, University of Southampton, Highfield, Southampton SO17 1BJ, UK\\
$^{7}$International Centre for Radio Astronomy Research--Curtin University, GPO Box U1987, Perth, WA 6845, Australia\\
$^{8}$New York University Abu Dhabi, P.O. Box 129188, Abu Dhabi, UAE\\
}

\date{Accepted XXX. Received YYY; in original form ZZZ}

\pubyear{2018}

\begin{document}
\label{firstpage}
\pagerange{\pageref{firstpage}--\pageref{lastpage}}
\maketitle

\begin{abstract}
Swift J1753.5-0127 (J1753) is a candidate black hole low-mass X-ray binary (BH-LMXB) that was discovered in outburst in May 2005. It remained in outburst for $\sim12$ years, exhibiting a wide range of variability on various timescales, before entering quiescence after two short-lived, low-luminosity ``mini-outbursts'' in April 2017. The unusually long outburst duration in such a short-period ($P_{\rm orb}\approx3.24$ hrs) source, and complex variability observed during this outburst period, challenges the predictions of the widely accepted disc-instability model, which has been shown to broadly reproduce the behaviour of LMXB systems well. The long-term behaviour observed in J1753 is reminiscent of the Z Cam class of dwarf novae, whereby variable mass transfer from the companion star drives unusual outbursts, characterized by stalled decays and abrupt changes in luminosity. Using sophisticated modelling of the multi-wavelength light curves and spectra of J1753, during the $\sim12$ years the source was active, we investigate the hypothesis that periods of enhanced mass transfer from the companion star may have driven this unusually long outburst. Our modelling suggests that J1753 is in fact a BH-LMXB analogue to Z Cam systems, where the variable mass transfer from the companion star is driven by the changing irradiation properties of the system, affecting both the disc and companion star.
\end{abstract}




\section{Introduction}

Low-mass X-ray binaries (LMXBs) are binary systems consisting of a compact object (black hole or neutron star) accreting matter from a low-mass ($\lesssim1M_{\odot}$) main sequence companion. Most LMXBs exhibit transient behaviour, characterized by long periods of quiescence (years to decades), followed by bright outbursts in which the X-ray and optical luminosity increases by several orders of magnitude \citep{chen97,tetarenko2016}.

The mechanism behind LMXB outbursts can broadly be explained with the disc-instability model \citep[DIM;][]{osaki1974,meyer81,smak83,smak84,cannizzo1985,cannizzo1993,huang89}. Originally developed to explain outbursts in dwarf novae, the DIM explains transient behaviour in the context of the accretion disc cycling between a hot, ionized outburst state and a cool, neutral, quiescent state. This limit-cycle is triggered by the accumulation of matter in the disc, eventually heating the disc until a substantial portion is ionized. In this hot, ionized state, the viscosity (i.e. ability to move angular momentum outwards) of the disc dramatically increases, ultimately resulting in matter rapidly falling onto the compact object, triggering a bright outburst.

The DIM has been shown to reproduce the global behaviour of a number of transient and persistent LMXBs \citep[e.g.][]{coriat2012,tetarenko2016}. However, to do so it must be modified to include irradiation heating from the inner accretion disc \citep{dubus1999,dubus2001}.
Most of the ultraviolet, optical and infrared light emitted by LMXB accretion discs is the result of reprocessed X-rays from the inner regions of the accretion flow heating the outer disc\footnote{Note that there exists alternative models attributing the dominant source of optical/UV emission (during LMXB outbursts) to self-produced synchrotron emission from a hot corona flow, rather than the irradiated disc itself. See \cite{Veledina-2017} and references therein for details.} \citep{JvP83,vanp1994,vanp96}. This X-ray irradiation is the dominant factor determining the temperature over the majority of the accretion disc during outburst. However, despite decades of theoretical work, the actual fraction of the central X-ray flux that is intercepted and reprocessed in the outer disc regions of LMXBs remains largely uncertain.

Outburst durations in LMXBs harbouring stellar-mass black holes (BH-LMXBs) are typically consistent with the viscous timescale of accretion discs \citep[$\sim$months, see e.g.][]{tetarenko2016}. Therefore, studying these binary systems during outburst allows us to probe the fundamental physics of accreting X-ray irradiated discs on accessible timescales. Recently, \citet{tetarenko2018} developed an analytical methodology to describe the outburst light curves of the Galactic BH-LMXB population in the context of the irradiated DIM. With this methodology, \citet{tetarenko2018} were able to quantify both angular-momentum and mass transport (via the $\alpha$-viscosity parameter; \citealt{ss73}) within these discs, and the physical properties of their X-ray irradiation heating. 

In another recent study, \citet{tetarenko2018b} demonstrated how the extremely diverse light curve morphology observed across the BH-LMXB population provides evidence for a likely temporally and spatially varying irradiation source heating the discs in these systems. In this paper, we build on their progress, presenting an alternative method to tackle this complex, multi-scale problem using the unusual BH-LMXB Swift J1753.5-0127 as a case study.

Swift J1753.5-0127 (hereafter J1753) was discovered by the Burst Alert Telescope \citep[BAT;][]{Barthelmy-2005} on board the {\em Neil Gehrels Swift Observatory} ({\em Swift}) in 2005 \citep{Palmer-2005}. The source luminosity peaked within a week, at a flux of $\sim$200 mCrab, as observed by the {\em Rossi X-Ray Timing Explorer}'s ({\em RXTE}) All-Sky Monitor \citep{CadolleBel-2007}. Identification of a $R\sim16$ optical counterpart \citep{Halpern-2005}, $\sim5$ magnitudes brighter than the limit of the Digitized Sky Survey, established J1753 as a new LMXB. 

Although the mass of the primary has not yet been dynamically measured, J1753 is classified as a BH candidate (BHC). {\em RXTE} observations revealed a 0.6 Hz quasi-periodic oscillation (QPO) with characteristics typical of BHCs \citep{Morgan-2005}. In addition, the hard X-ray spectrum was found to exhibit a hard power-law tail up to $\sim600$ keV, very typical of a BHC in a hard accretion state \citep{CadolleBel-2007}. By investigating the double-peaked hydrogen recombination lines in the optical spectrum, \citet{Shaw-2016d} calculated a conservative lower limit to the mass of the primary, $M_1>7.4\pm1.2{\rm M}_{\odot}$, strongly favouring a BH nature for the compact object in J1753. The orbital period of J1753 is known to be $P_{\rm orb}\approx3.24$h\footnote{This photometric period is likely a superhump period, slightly larger than the orbital period, due to an extreme mass ratio between the primary and the secondary} from periodic variability in its optical light curves observed during outburst \citep{zur8}. The high amplitude variations in the $R$-band orbital light curve suggest that the inclination, $i$, is high. However, the source is not dipping nor eclipsing, and a measurement of $i$ remains elusive \citep[see][]{Shaw-2016d}, so previous studies have adopted values of $i\gtrsim40^{\circ}$ \citep[e.g.][]{Neustroev-2014,tomsick2015}. The distance, $D$, to J1753 is also an unknown, with no measurement from radio parallax, and a poorly constrained optical parallax from {\em Gaia} \citep{Gandhi-2018}. However, \citet{zur8} estimate $D>7.2$ kpc based on the relationship between $P_{\rm orb}$ and the observed peak $V$-band magnitude \citep{Shahbaz-1998a}. \citet{Froning-2014}, on the other hand, derive $D<7.7$ kpc from modelling the UV spectrum. \cite{Gandhi-2018}, using prior knowledge of the density of LMXBs in the bulge, disc, and sphere of the Milky Way Galaxy and Bayesian statistical techniques, estimate a posterior distribution on distance of $D=7.74_{-2.13}^{+5.00}$ kpc. We make use of this distance estimate in this paper. 

Almost immediately after its peak, the flux of J1753 began to decline exponentially, as is typical with similar BH-LMXBs. However, instead of returning to quiescence the decay stalled, remaining at roughly constant flux at approximately $\sim20$ mCrab for over 6 months \citep{CadolleBel-2007} before appearing to increase in flux once more \citep[see e.g.][]{Shaw-2013a}. For the following $\sim11$ years, the source continued to be active, exhibiting significant long term variability \citep[see e.g.][]{Soleri-2013,Shaw-2013a}. J1753 remained a persistent LMXB in a hard accretion state for the majority of this prolonged period of activity, aside from a brief excursion to a low-luminosity soft, disc-dominated accretion state in 2015 \citep{Shaw-2016b,Rushton-2016}.

In November 2016, observations of J1753 with the Faulkes Telescope North indicated a fading optical flux \citep{Russell-2016,AlQasim-2016}. \citet{Shaw-2016c} confirmed the decay with a non-detection by both the {\em Swift} X-ray Telescope \citep[XRT;][]{Burrows-2005} and Ultra-Violet/Optical Telescope \citep[UVOT;][]{Roming-2005}. Unfortunately due to the source becoming Sun constrained soon after the decay to quiescence was noted, follow-up opportunities were limited \citep{Neustroev-2016,Plotkin-2016}. Upon emerging from the Sun constraint, J1753 was found to be active once more, having undergone a mini-outburst \citep{AlQasim-2017,Bright-2017,Tomsick-2017}\footnote{We choose the nomenclature `mini-outburst' to remain consistent with \citet{plotkin2017}, who note that the flux during this period was at similar levels prior to quiescence. Following \citet{chen97}, such behaviour is referred to as a `mini-outburst.' See discussion in Zhang et al. (2018, submitted).} which lasted from late January 2017 until April 2017, when the flux appeared to return to quiescent levels \citep{Shaw-2017a}. However, in late April 2017, J1753 underwent another, shorter duration mini-outburst \citep{Bernardini-2017}, before finally entering a seemingly more permanent quiescent state in July of the same year \citep{Zhang-2017}.

In this paper, we initially focus on the first mini-outburst of J1753 during its descent into quiescence. We utilize near-simultaneous near-infrared (NIR), optical, ultraviolet (UV) and X-ray data in an attempt to model and constrain how the physical properties of the X-ray irradiation source heating an accretion disc varies over time during outburst. To achieve this we have developed a Bayesian algorithm to fit the observed UV, optical and NIR (UVOIR) spectral energy distribution (SEDs) with an irradiated disc model. We then compare the mini-outburst with the main outburst of J1753 in an attempt to investigate the nature of its long-term behaviour.

The paper format is as follows: Section \ref{sec:modelling} describes in detail the irradiated disc model and the Bayesian algorithm used to fit such a model to observed SEDs. Section \ref{sec:obs} describes the observations we have utilized in this work. Section \ref{sec:results} presents the results from fitting the X-ray spectra, UVOIR SEDs, X-ray light curves, and UVOIR-X-ray correlations during the 12 yrs that J1753 was active. Section \ref{sec:discuss} discusses the nature of the unusual long-term behaviour displayed by J1753 over the past 12 years, drawing a parallel with Z Cam type dwarf novae. Lastly, Section \ref{sec:conclusions} provides a summary of this work.

\section{Modelling X-ray Irradiated Accretion Discs}
\label{sec:modelling}

\subsection{The Irradiated Disc Model}\label{sec:disc_model}

 Simultaneous, multi-wavelength data sets can constrain the evolution and X-ray irradiation and heating of the accretion disc through LMXB outbursts.
By modelling the observed UVOIR SED of a LMXB system, at different times throughout an outburst, one can effectively  track the 
evolution of the physical properties of the X-ray irradiation source heating the disc in the system (e.g. \citealt{hynes2005}, \citealt{russell2006}, \citealt{meshcheryakov2018}). 

Starting with a power-law temperature distribution of the form,
\begin{equation}
T(R)=T_{\rm in}\left( \frac{R}{R_{\rm in}} \right)^{-n},
\label{eq:r^n}
\end{equation}
and assuming a local blackbody spectrum in a particular disc annulus at radius $R$, we can describe the integrated SED of the disc in terms of 3 parameters: the inner disc temperature ($T_{\rm in}$), temperature of the outer disc ($T_{\rm out}$), and the disc normalization ($N_{\rm disc}$). The resulting SED takes the form,
\begin{equation}
F_{\nu}=
N_{\rm disc}^2  T_{\rm in,K}^{\frac{2}{n}}
\nu^{3-\frac{2}{n}} I(x),
\label{eq:flux_bbi2}
\end{equation}
where $I(x)$ is Planck's law integrated over the disc, using the general substitution $x=h\nu/k_b T$, and must be evaluated numerically. The normalization takes the form
\begin{equation}
N_{\rm disc}=\left(\frac{4 \pi h^{1-\frac{2}{n}} k_b^{\frac{2}{n}}\cos(i)}{n c^2} \right)^{1/2}\left(\frac{R_{\rm in}}{\rm{cm}} \right) \left(\frac{D}{\rm{cm}} \right)^{-1}.
\end{equation}

To model the X-ray irradiation heating LMXB accretion discs, we use the prescription presented in \cite{dubus2001},
 \begin{equation}
T_{\rm out}^4=\frac{C_{\rm irr} L_X}{4 \pi \sigma_{\rm SB}  R^2},
\label{eq:irr}
\end{equation}
where the irradiating flux drops off with the inverse square of the disc radius. Here $C_{\rm irr}$ is a constant representing the fraction of the X-ray luminosity intercepted and reprocessed by the disc.

We consider temperature profiles (Eq. \ref{eq:r^n}) with the power-law index $n$ equal to $1/2$ (i.e., standard irradiated disc) or $3/7$ (i.e., a maximally-irradiated, isothermal disc).
In addition to these irradiated cases, we also consider the $n=3/4$ model,  appropriate for a non-irradiated disc. 
See \cite{hynes2002} and \cite{hynes2005} for detailed discussions of the various temperature profile options to fit UVOIR SEDs of LMXBs.

In addition to reprocessed X-rays, OIR emission may also be produced by relativistic jets, as seen to dominate during the hard accretion states of some BH-LMXBs (e.g., see \citealt{homan2005,russell2006}).
These jets produce a spectrum consisting of a flat, optically thick component and a steep, optically thin component that can extend from the radio to OIR wavelengths \citep{fender2001,corbel2002,homan2005,russell2006,chaty2011,rahoui2015}. In the UVOIR wavelength range, we model only the optically thin part of the jet spectrum, as the jet break in BH-LMXBs is usually at lower frequencies \citep[see e.g.][]{Gandhi-2011}. Thus, when required, we add a single power-law function to the disc model described above. The power law takes form $F_{\nu}=N_{\rm pl}\nu^{-\beta}$, where $N_{\rm pl}$ is the power-law normalization and $\beta$ is the power-law index.

\subsection{Markov-Chain Monte Carlo (MCMC) Algorithm}\label{sec:mcmc_algorithm}

We use a Bayesian algorithm to estimate the three to five parameters required to describe the observed UVOIR SEDs of a BH-LMXB in outburst: the (i) inner disc temperature ($T_{\rm in}$), (ii) temperature of the outer disc ($T_{\rm out}$), (iii) disc normalisation ($N_{\rm disc}$), (iv) power-law normalisation ($N_{\rm pl}$) and (v) power-law index ($\beta$). Specifically, we make use of the {\tt emcee} {\sc python} package \citep{for2013}, an implementation of Goodman \& Weare's Affine Invariant MCMC Ensemble Sampler \citep{goodw10}, to fit the observed UVOIR SEDs with the models described in Section \ref{sec:disc_model}. 


{\tt emcee} works by using a modified version of the Metropolis-Hastings Algorithm, where an ensemble of ``walkers'' simultaneously explores the parameter space. We use a number of walkers equal to 10 times the model dimensions to fit our SEDs. 
For the initial inspection of our parameter space, we use a harmony search global optimization algorithm called {\tt pyHarmonySearch} \citep{geem2001} for all parameters except for $T_{\rm in}$. 
The ``best guess'' provided by {\tt pyHarmonySearch} acts as a starting point for the MCMC walkers.

The prior distribution and initialization for $T_{\rm in}$ is set using the constraints from the X-ray spectral fits. As no disc component is observed in the $0.6-10$ keV band XRT spectra, we use a uniform distribution ($0.0<T_{\rm in} {\rm (keV)}<0.6$) as the prior for $T_{\rm in}$. The prior distributions for the remaining four parameters ($T_{\rm out}$, $N_{\rm disc}$, $N_{\rm pl}$, $\beta$) are chosen to be Gaussians, with means set using the {\tt pyHarmonySearch} harmony search global optimization algorithm. 

After initialization, the MCMC is run on each SED, starting with a ``burn-in'' phase where an ensemble of ``walkers'' are evolved over a series of 500 steps. 
Following the burn-in phase, the MCMC is run again, until convergence. 
The MCMC algorithm outputs the converged solution in the form of posterior distributions of each parameter. We take the median and 1$\sigma$ (68\%) confidence interval of each posterior distribution as the best-fit result for each parameter.

\section{Observations and Data Reduction}
\label{sec:obs}

\subsection{X-ray}

\subsubsection{Swift/XRT}\label{sec:swift_data}

We utilised 17 {\em Swift}/XRT observations of J1753 during the two observed mini-outbursts (ObsIDs 00030090116 -- 00030090137), from 2017 Feb 16 to May 15. XRT observed in auto exposure mode for the majority of this time, adjusting the CCD readout mode between windowed timing (WT) and photon counting (PC) modes depending on the source count rate.

Data were reprocessed using the {\sc heasoft} v6.23 task {\sc xrtpipeline}. WT mode observations were extracted from a 20 pixel ($\approx47''$) radius circular aperture centered on the source. Background spectra in WT mode were extracted from an annulus centered on the source with inner and outer radii of 80 and 120 pixels, respectively.\footnote{http://www.swift.ac.uk/analysis/xrt/backscal.php} PC mode observations are more susceptible to photon pile-up than WT mode observations. Thus, PC mode source spectra were initially extracted from a circular region of the same radius as in WT mode, and the average count rate was then calculated in order to determine if photon pile-up was significant. Observations with count rates higher than $0.5$ count s$^{-1}$ were re-extracted using an annular region with an outer radius of 20 pixels and the central portion of the point spread function excluded. The radius of the excluded region was determined by NASA's {\sc ximage} package\footnote{http://www.swift.ac.uk/analysis/xrt/pileup.php} and ranged from $\sim2-4$ pixels. PC mode background spectra were extracted from an annulus centered on the source with inner and outer radii of 50 and 70 pixels, respectively.

Source and background spectra were extracted using the {\sc heasoft} tool {\sc xrtproducts} and were grouped such that each energy bin contained a minimum of one count. Ancillary response files were generated with {\sc xrtmkarf} and the relevant canned response matrix files were obtained from the High Energy Astrophysics Science Archive Research Center (HEASARC) calibration database (CALDB). All X-ray spectral fits were performed in {\sc xspec} v.12.10.0 \citep{Arnaud-1996}, using Cash statistics for background subtracted spectra \citep[W-statistic;][]{Cash-1979}. Interstellar absorption was accounted for with the {\tt tbabs} model with \citet{Wilms-2000} abundances and \citet{Verner-1996} photoionisation cross-sections. Unabsorbed 2--10 keV fluxes were calculated using the {\tt cflux} model. All uncertainties on parameters derived from the X-ray spectral fits are at the 90\% confidence level, unless otherwise stated.

For observations where the source was not detected by {\em Swift}/XRT, we derived 90\% confidence upper limits on the 2--10 keV flux using the methods described by \citet{Gehrels-1986}. We assumed a power-law model with index $\Gamma=1.7$ \citep[e.g.][]{Shaw-2016a} and a hydrogen column density $N_{\rm H}=2\times10^{21}$ cm$^{-2}$ \citep{Froning-2014}. For observations where a source was clearly detected (at $>99$\% confidence), but at a low count rate such that X-ray spectral fitting was not possible, we followed the procedure of \citet{plotkin2017}. Fluxes in this case were derived assuming the same model as above and we derived uncertainties using 90\% confidence intervals from \citet{Kraft-1991}, factoring in a photon index that was allowed vary from $1 < \Gamma < 2.5$.



\subsubsection{Long-term X-ray Light Curves}\label{sec:longterm_data}

We  built a long-term X-ray light curve for J1753 using  data from the (i) {\em RXTE}/Proportional Counter Array (PCA) and (ii) {\em Swift}/XRT.

We collected all the {\em RXTE}/PCA data from the WATCHDOG project \citep{tetarenko2016}. These datasets include (i) good pointed PCA observations (i.e. no scans or slews) available in the HEASARC archive over the entire {\em RXTE} mission. We collected all available {\em Swift}/XRT data (including WT and PC mode pointed observations) between May 2005 and April 2017 using the {\em Swift}/XRT online product builder\footnote{http://www.swift.ac.uk/user\_objects/index.php}\citep{evans2009}.

All light curves were collected in the 2--10 keV band. Individual instrument count rates were converted into flux using Crabs as a baseline unit and calculating approximate count rate equivalences. See \citet{tetarenko2016} for details on this method.

 \begin{figure}
  \center
\includegraphics[width=\columnwidth]{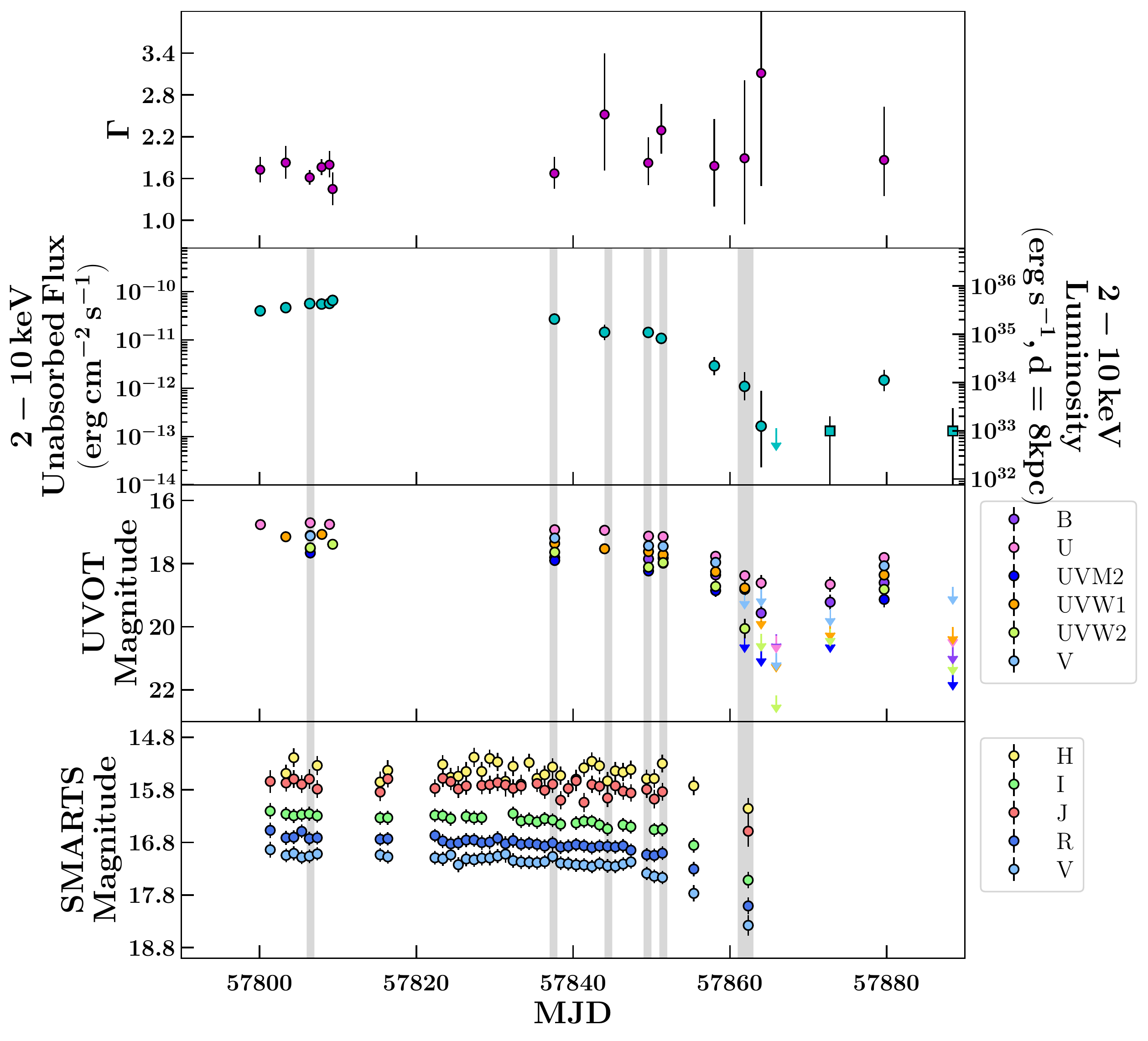}
\caption{Multi-wavelength view of the J1753 mini-outbursts. Our six individual epochs with simultaneous/quasi-simultaneous UVOIR data  are indicated with grey vertical bars. The panels give the photon index from X-ray spectral fits (top panel), and the {\em Swift}/XRT (second panel), {\em Swift}/UVOT (third panel), and {\em SMARTS} (bottom panel) lightcurves. Upper limits are displayed with coloured arrows and low count rate flux estimates are represented by square markers (see Section \ref{sec:swift_data} for details). Note that the UVOT and {\em SMARTS} data displayed here are not dereddened. 
}
\label{fig:stacked_lc}
\end{figure}

\subsection{UVOIR}

For constructing and fitting the SEDs, the UVOIR photometry described in the following sections is corrected for interstellar extinction according to \citet{Fitzpatrick-1999}. To deredden the data we use the {\tt specutils} package in {\sc python}, and $E(B-V)=0.45$ \citep{Froning-2014}.

\subsubsection{Swift/UVOT}\label{sec:uvot_data}

{\em Swift}/UVOT is a 30cm diameter telescope which operates simultaneously with the XRT. When possible, we obtained observations in each of the six filters available for UVOT, from the UV ($UVW2, UVM2, UVW1$) to optical ($U, B, V$). As a result, we have 17 UVOT observations simultaneous with the XRT exposures described in Section \ref{sec:swift_data}.

Aperture photometry was performed on the images with the {\sc heasoft} tool {\sc uvotsource}, using a $5''$ radius circular region centered on the source. The background was measured using a $20''$ radius circular aperture centred on a source free region. All reported magnitudes are in the Vega system and have been converted into flux densities using the known flux zero-points for each filter. Uncertainties include the statistical $1\sigma$ error and a systematic uncertainty that accounts for the uncertainty in the shape of the UVOT point spread function (PSF).

\subsubsection{SMARTS}\label{sec:smarts_data}

Upon the discovery of the first mini-outburst, we commenced optical and NIR monitoring of J1753 with A Novel Dual Imaging CAMera \citep[ANDICAM;][]{Depoy-2003} on the 1.3m {\em Small \& Moderate Aperture Research Telescope System} \citep[{\em SMARTS};][]{Subasavage-2010} at Cerro Tololo, Chile. An observing sequence consisted of observations in $V, R, I, J, H$ and $K$ bands\footnote{The $K$-band images were of poor quality due to the rapidly changing sky background at longer wavelengths, so we choose not to include them in this work.} with exposure times of 360 s in the optical filters and 30 s in each of 8 to 10 dithered images in the NIR filters. Images were reduced using the Image Reduction and Analysis Facility \citep[IRAF;][]{Tody-1986,Tody-1993}, following the standard procedures described by \citet{Buxton-2012}. Point-spread function photometry was performed on all images. The magnitudes were calibrated to the Vega system with respect to nearby stars in the field, with absolute calibration performed on clear nights using \citet{Landolt-1992} standards and the 2 Micron All-Sky Survey (2MASS) catalogue \citep{Skrutskie-2006} in the optical and NIR, respectively. 

\subsection{2014 observations}\label{sec:2014_obs}

We also utilise multi-wavelength data from 2014 Apr 2 -- 5 observations of J1753, taken with an array of instruments. In UVOIR, we utilise observations performed in all six {\em Swift}/UVOT filters, in the $g', r', i'$ and $z'$ bands using the Lulin 41cm {\em Super-Light Telescope} ({\em SLT}), located in Taiwan and in the $B, V, J$ and $K_s$ bands using the HONIR instrument on the 1.5m {\em Kanata} telescope at the Higashi-Hiroshima Observatory, Japan. We also utilise a single 2.4ks {\em Swift}/XRT observation performed on 2014 Apr 5 (ObsID 00080730001). The data reduction of all of the 2014 data is described by \citet{tomsick2015}.

\subsection{The Bolometric Correction}\label{sec:bolcorr}

To convert the 2--10 keV unabsorbed flux from {\em Swift}/XRT and {\em RXTE}/PCA, to a bolometric flux, we use a combination of the (i) bolometric corrections estimated for each accretion state by \citet{migliari2006}, (ii) WATCHDOG project's online Accretion-State-By-Day tool\footnote{http://astro.physics.ualberta.ca/WATCHDOG}, which provides accretion state information on daily timescales during outbursts of BH-LMXBs, and (iii) \cite{Shaw-2016b}, who quantify J1753's brief excursion to the soft state in 2015.

We note that using this method assumes that the bolometric conversion factor remains constant when the source is in each individual accretion state. This may not always be the case. Specifically for J1753, the high-energy cutoff in the X-ray spectrum has been observed to vary between $\sim100-600$ keV throughout its 12 years of activity, where it mostly remained in a hard accretion state (see e.g., \citealt{CadolleBel-2007,kajava2016}). Without any continuous X-ray monitoring at $>$100 keV we have no way of quantifying whether (or how) this correction may change throughout the main outburst, standstill period, or mini-outbursts. Thus, we 
assume that the bolometric correction remains constant and acknowledge that if this is not the case, some uncertainty will be introduced into our calculation of bolometric flux.

\section{Results}
\label{sec:results}
\subsection{X-ray Spectral Fitting}\label{sec:xspec_fits}

All X-ray spectra are adequately fit with an absorbed power-law model ({\tt tbabs*powerlaw} in {\sc xspec} syntax), with no model fit improved by the addition of a {\tt diskbb} model. The best-fit power-law indices ($\Gamma$) and unabsorbed 2--10 keV fluxes at each {\em Swift} epoch are plotted in Fig. \ref{fig:stacked_lc}. For the duration of the two mini-outbursts shown in Fig. \ref{fig:stacked_lc}, $\Gamma$ remains roughly constant at $\sim1.6$, a value consistent with that typically expected of an LMXB in a hard accretion state. However, whilst $\Gamma$ remains roughly constant, the flux does not, first rising and then decaying as the first mini-outburst progresses. We also note the onset of the second mini-outburst at MJD $\sim57880$, which was  investigated by \citet{plotkin2017}.

\begin{table*}
{
\setlength{\tabcolsep}{3pt}
\centering
\caption{ MCMC SED Fitting Results}
\medskip
\label{tab:outfits}
\begin{tabular}{lcccccccccc} 
\hline
Epoch & Date & Function&$T_{\rm in}^{a}$&$T_{\rm out}$&$N_{\rm disc}$ $^b$&$N_{\rm pl}$ $^c$&$\beta$&$\log_{10}(R_{\rm out}/R_{\rm in})$&$F_{\rm disc} \times 10^{-10}$&$R_{\rm in}$ $^d$ \\
&(MJD) & Type&(keV)&($\times 10^{3}$ K)&&&&&($\rm{erg\, cm^{-2} \,s^{-1}}$)&($\times R_g$)\\[0.045cm]
  \hline
  \multicolumn{11}{c}{Standard Irradiated Disc ($n=1/2$)}\\ \hline
1 & 57806-57807 & disc+pl&$0.19_{-0.02}^{+0.03}$&$11.0\pm0.5$&$5.0_{-1.0}^{+1.1}$&$3.6_{-1.1}^{+0.9}$&$1.18_{-0.01}^{+0.02}$&$4.6\pm0.1$&$1.9_{-1.2}^{+3.8}$&$2.1_{-1.7}^{+2.5}$ \\[0.2cm]
2 & 57837-57838 & disc+pl&$0.16_{-0.02}^{+0.04}$&$9.7\pm0.5$&$5.9_{-1.1}^{+1.2}$&$9.3_{-3.1}^{+2.9}$&$1.21_{-0.02}^{+0.03}$&$4.56_{-0.04}^{+0.13}$&$1.4_{-0.9}^{+3.4}$&$2.4_{-1.9}^{+3.0}$ \\[0.2cm]
3 & 57844 & disc &$0.10_{-0.01}^{+0.05}$&$6.2_{-0.3}^{+0.4}$&$12\pm1$&$<11$&\nodata&$4.50_{-0.02}^{+0.30}$&$0.8_{-0.3}^{+4.2}$&$5.0_{-4.6}^{+5.5}$
\\[0.2cm]
4 & 57849 & disc &$0.090_{-0.004}^{+0.055}$&$6.2_{-0.4}^{+0.3}$&$14\pm1$&$<3.1$&\nodata&$4.48_{-0.03}^{+0.32}$&$0.7_{-0.2}^{+4.8}$&$5.7_{-5.2}^{+6.1}$ \\[0.2cm]
5 & 57851 & disc+pl &$0.04_{-0.01}^{+0.02}$&$8.1_{-0.9}^{+1.2}$&$62_{-27}^{+15}$&$4.7_{-2.8}^{+3.5}$&$1.17_{-0.03}^{+0.02}$&$3.58_{-0.04}^{+0.21}$&$0.6_{-0.6}^{+4.5}$ &$26_{-14}^{+32}$\\[0.2cm]
6 & 57861-57862 & disc+pl &$0.03\pm0.01$&$6.1_{-0.8}^{+0.6}$&$61\pm2$&$132\pm3$&$1.31_{-0.07}^{+0.05}$&$3.5\pm0.1$&$0.2_{-0.1}^{+0.2}$&$25_{-25}^{+26}$ \\[0.2cm]
*& 56752 &disc&$0.06\pm0.01$&$6.8\pm0.4$&$46\pm15$&\nodata&\nodata&$4.0\pm0.1$&$1.1_{-0.8}^{+3.5}$&$19_{-13}^{+26}$\\[0.045cm]
\hline
  \multicolumn{11}{c}{Fully Irradiated Disc ($n=3/7$)}\\ \hline
1 & 57806-57807 & disc+pl&$0.09_{-0.01}^{+0.02}$&$12.5\pm0.6$&$1.6_{-0.5}^{+0.6}$&$1.9\pm1.0$&$1.15_{-0.02}^{+0.01}$&$4.5\pm0.1$&$0.8_{-0.6}^{+2.4}$&$2.3_{-1.5}^{+3.2}$ \\[0.2cm] 
2 & 57837-57838 & disc+pl&$0.09\pm0.01$&$11.8\pm0.6$&$1.7_{-0.3}^{+0.4}$&$1.2\pm0.5$&$1.15_{-0.02}^{+0.01}$&$4.50_{-0.03}^{+0.04}$&$0.7_{-0.4}^{+0.7}$&$2.3_{-1.9}^{+2.8}$ \\[0.2cm]
3 & 57844 & disc &$0.04_{-0.01}^{+0.02}$&$7.9_{-0.7}^{+0.6}$&$6.7_{-2.1}^{+1.3}$&$<6.5$&\nodata&$4.17_{-0.05}^{+0.26}$&$0.5_{-0.3}^{+2.7}$&$9.4_{-6.5}^{+11.3}$ \\[0.2cm]
4 & 57849 & disc &$0.035_{-0.003}^{+0.004}$&$6.6\pm0.1$&$7.8_{-1.6}^{+1.8}$&$<2.4$&\nodata&$4.2\pm0.1$&$0.3_{-0.2}^{+0.5}$&$11_{-9}^{+13}$ \\[0.2cm]
5 & 57851 & disc+pl &$0.034_{-0.002}^{+0.003}$&$11.3_{-0.7}^{+0.6}$&$13\pm2$&$0.8\pm0.3$&$1.12_{-0.02}^{+0.01}$&$3.590_{-0.003}^{+0.040}$&$0.5_{-0.2}^{+0.4}$ &$18_{-15}^{+20}$\\[0.2cm]
6 & 57861-57862 & disc+pl &$0.016_{-0.002}^{+0.001}$&$6.2\pm0.4$&$24\pm4$&$6.5_{-2.4}^{+2.9}$&$1.20_{-0.09}^{+0.08}$&$3.420_{-0.004}^{+0.002}$&$0.07_{-0.03}^{+0.05}$&$34_{-29}^{+40}$ \\[0.2cm]
*& 56752 &disc&$0.04\pm0.01$&$7.3\pm0.4$&$9.0_{-3.8}^{+3.4}$&\nodata&\nodata&$4.2_{-0.1}^{+0.2}$&$0.6_{-0.5}^{+2.7}$&$13_{-7}^{+18}$\\[0.045cm]
  \hline
  \multicolumn{11}{c}{Non-Irradiated Disc ($n=3/4$)}\\ \hline
 1 & 57806-57807 & disc+pl&$(5.2_{-0.1}^{+0.2})\times10^{-3}$&$6.1\pm0.7$&$1.6\pm0.1$&$3.5_{-0.3}^{+0.7}$&$1.54_{-0.06}^{+0.04}$&$1.3\pm0.1$&$0.9\pm0.2$& $(6.2_{-5.9}^{+6.4})\times 10^{3}$ \\[0.2cm] 

 2 & 57837-57838 &
 disc+pl&$(4.3\pm0.1)\times10^{-3}$&$6.1_{-0.9}^{+1.0}$&$1.9\pm0.1$&$0.030\pm0.001$&$1.40_{-0.01}^{+0.02}$&$1.2\pm0.1$&$0.6_{-0.1}^{+0.2}$& $(7.3_{-7.0}^{+7.6})\times 10^{3}$ \\[0.2cm]
 
 3 & 57844 & disc &$(2.9\pm0.1)\times10^{-3}$&$4.4\pm0.6$&$3.3\pm0.1$&$<0.06$&\nodata&$1.2\pm0.1$&$0.34_{-0.05}^{+0.04}$&  $(1.3_{-1.2}^{+1.3})\times 10^{4}$ \\[0.2cm]

 4 & 57849 & disc &$(2.60\pm0.05)\times10^{-3}$&$4.1\pm0.6$&$3.8\pm0.1$&$<464$&\nodata&$1.2\pm0.1$&$0.30_{-0.03}^{+0.04}$&$(1.5_{-1.4}^{+1.5})\times 10^{4}$ \\[0.2cm]
 5 & 57851 & disc+pl &$(2.10\pm0.04)\times10^{-3}$&$8.0\pm0.8$&$5.7\pm0.1$&$400\pm10$&$1.67\pm0.01$&$0.66_{-0.04}^{+0.05}$&$0.26_{-0.02}^{+0.02}$ &$(2.2\pm2.2)\times 10^{4}$\\[0.2cm]
 6 & 57861-57862 & disc+pl &$(1.3\pm0.1)\times10^{-3}$&$5.7_{-0.9}^{+0.8}$&$7.8\pm0.7$&$5505_{-107}^{+99}$&$1.82_{-0.04}^{+0.13}$&$0.57_{-0.05}^{+0.06}$&$0.06_{-0.06}^{+0.02}$&$(3.0_{-2.7}^{+3.2})\times 10^{4}$
 \\[0.2cm] 
 *& 56752 &disc&$(2.3\pm0.2)\times10^{-3}$&$5.9\pm0.4$&$6.4_{-0.7}^{+0.6}$&\nodata&\nodata&$0.880_{-0.005}^{+0.012}$&$0.5_{-0.2}^{+0.4}$&$(2.4_{-2.2}^{+2.7})\times 10^{4}$\\[0.045cm] \hline
    \multicolumn{11}{p{1.5\columnwidth}}{\hangindent=1ex $^a$$1{\rm keV}\approx1.2\times10^{7}{\rm K}$}.\\
    \multicolumn{11}{p{1.5\columnwidth}}{\hangindent=1ex $^b$The disc normalization $N_{\rm disc}$: $\times10^{-19}$ ($n=1/2$), $\times10^{-15}$ ($n=3/7$), and $\times10^{-16}$ ($n=3/4$).}\\
    \multicolumn{11}{p{2.1\columnwidth}}{\hangindent=1ex $^c$The PL normalization $N_{\rm pl}$: $\times10^{19}$ ($n=1/2$ and $n=3/7$) and $\times10^{24}$ ($n=3/4$). Upper limits on $N_{\rm pl}$ are found using the disc flux in the H-band and assuming the $\beta$ of the closest available observation.}\\
    \multicolumn{11}{p{1.5\columnwidth}}{\hangindent=1ex $^d$Calculated for an assumed distance of $D=8$ kpc and inclination  $i=40^{\circ}$.}\\
    \multicolumn{11}{p{1.5\columnwidth}}{\hangindent=1ex $^*$Best fit to data taken during the main outburst \citep{tomsick2015}. See Sections \ref{sec:2014_obs} and \ref{sec:discuss} for details.}\\
\end{tabular}\\
}
\end{table*}

\subsection{UVOIR SED Fitting}\label{sec:uvoir_fits}

We fit the UVOIR SEDs from six individual epochs during the J1753 mini-outburst (see Fig. \ref{fig:stacked_lc} and Table \ref{tab:outfits}) with the disc+power-law model described in Section \ref{sec:disc_model}. Each epoch has strictly simultaneous or quasi-simultaneous (within 1 day) UV through IR data from {\em Swift}/UVOT and {\em SMARTS} (see Sections \ref{sec:uvot_data} and \ref{sec:smarts_data}).

All broadband SEDs during each of our six epochs are presented in Fig. \ref{fig:broadband_seds}. Each SED has been plotted in frequency space with the (dereddened) data colour-coded by wavelength: UVOIR (UVOT and {\em SMARTS}; red), X-ray (XRT: $2-10$ keV; purple), and Radio (VLA: 9.8 GHz and AMI: 15.5 GHz from \citealt{plotkin2017}; yellow). Note that only the UVOIR data are fit with our MCMC algorithm. X-ray and radio data 
show the multi-wavelength behaviour of the source\footnote{To derive the X-ray fluxes plotted in Fig. \ref{fig:broadband_seds}, we fit the X-ray data with the {\tt pegpwrlw} model, where the model normalization is the flux density in $\mu{\rm Jy}$ at a specified energy. We derive this normalization at discrete energies in the range 1--10 keV and plot them in Fig. \ref{fig:broadband_seds}.}. All X-ray and radio data plotted are either simultaneous or quasi-simultaneous (within 1 day) with the UVOIR data, with the exception of Epoch 1, in which the closest available radio data were taken three days prior to the UVOIR data. 
Epochs 1,2,5, and 6 required the addition of a power-law component (indicative of the presence of emission from a jet) to the disc component to sufficiently match the UVOIR behaviour, while Epochs 3 and 4 were well fit with a pure disc model. 

\begin{figure}
  \center
\includegraphics[width=\columnwidth]{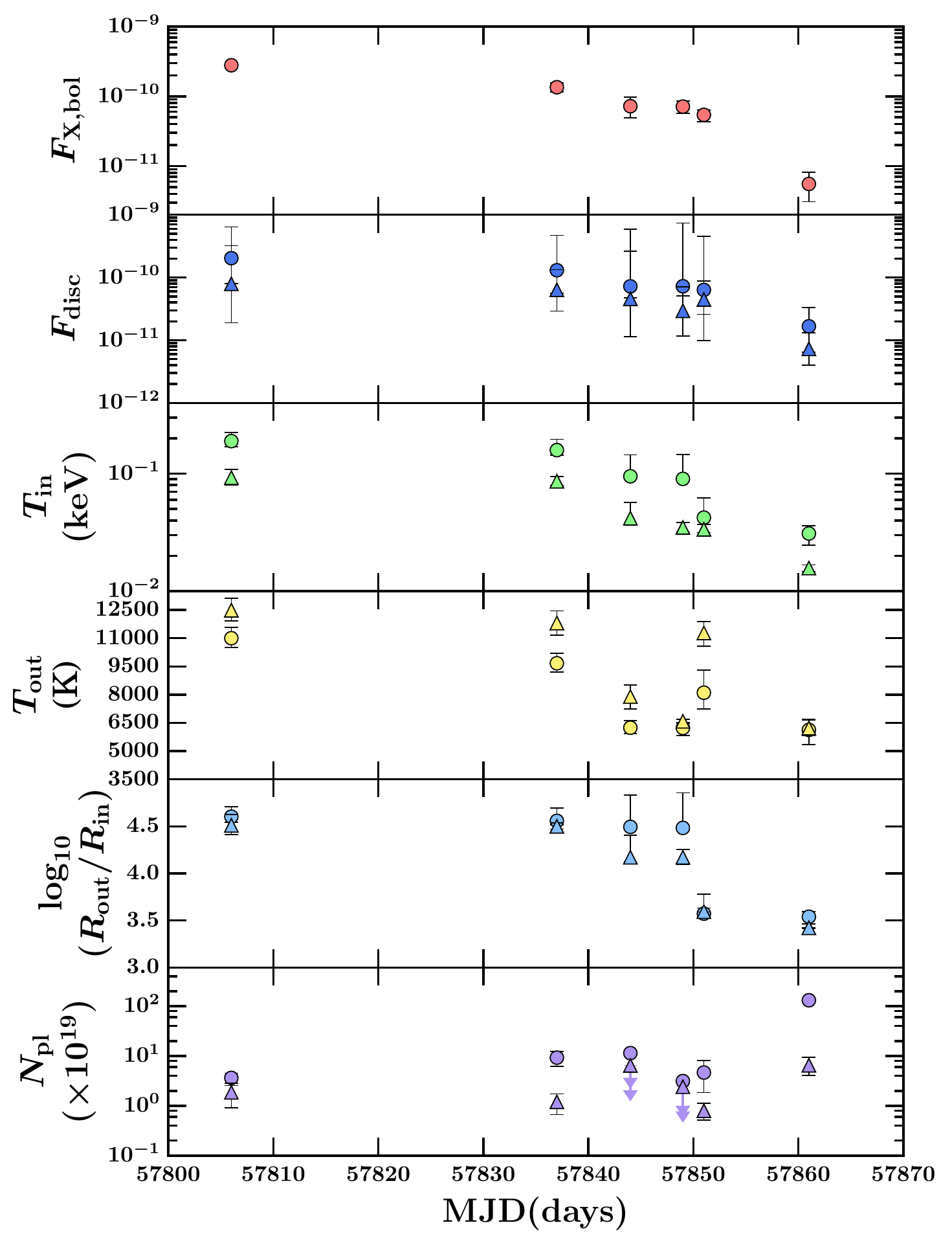}
\caption{Irradiated disc model fits to the J1753 mini-outburst. For each of our six epochs, we plot: bolometric X-ray flux (top panel), integrated irradiated disc flux (second panel), and the parameters derived from the MCMC algorithm: inner disc temperature $T_{\rm in}$ (third panel), outer disc temperature $T_{\rm out}$ (fourth panel), log of the ratio of outer to inner disc radius $\log_{10}(R_{\rm out}/R_{\rm in})$ (fifth panel), and power-law normalization $N_{\rm pl}$ (bottom panel). Filled circles ($n=1/2$) and triangles ($n=3/7$) are the results from fitting the SEDs with our irradiated disc model with two different temperature profiles, $T\propto R^{-n}$. The errors on bolometric flux include (90\% confidence) statistical instrument uncertainty only. 
Errors on fit parameters are $1\sigma$ confidence.
Upper limits on $N_{\rm pl}$ are displayed as purple arrows.
}
\label{fig:fit_params}
\end{figure}

\begin{figure}
  \center
\includegraphics[width=\columnwidth]{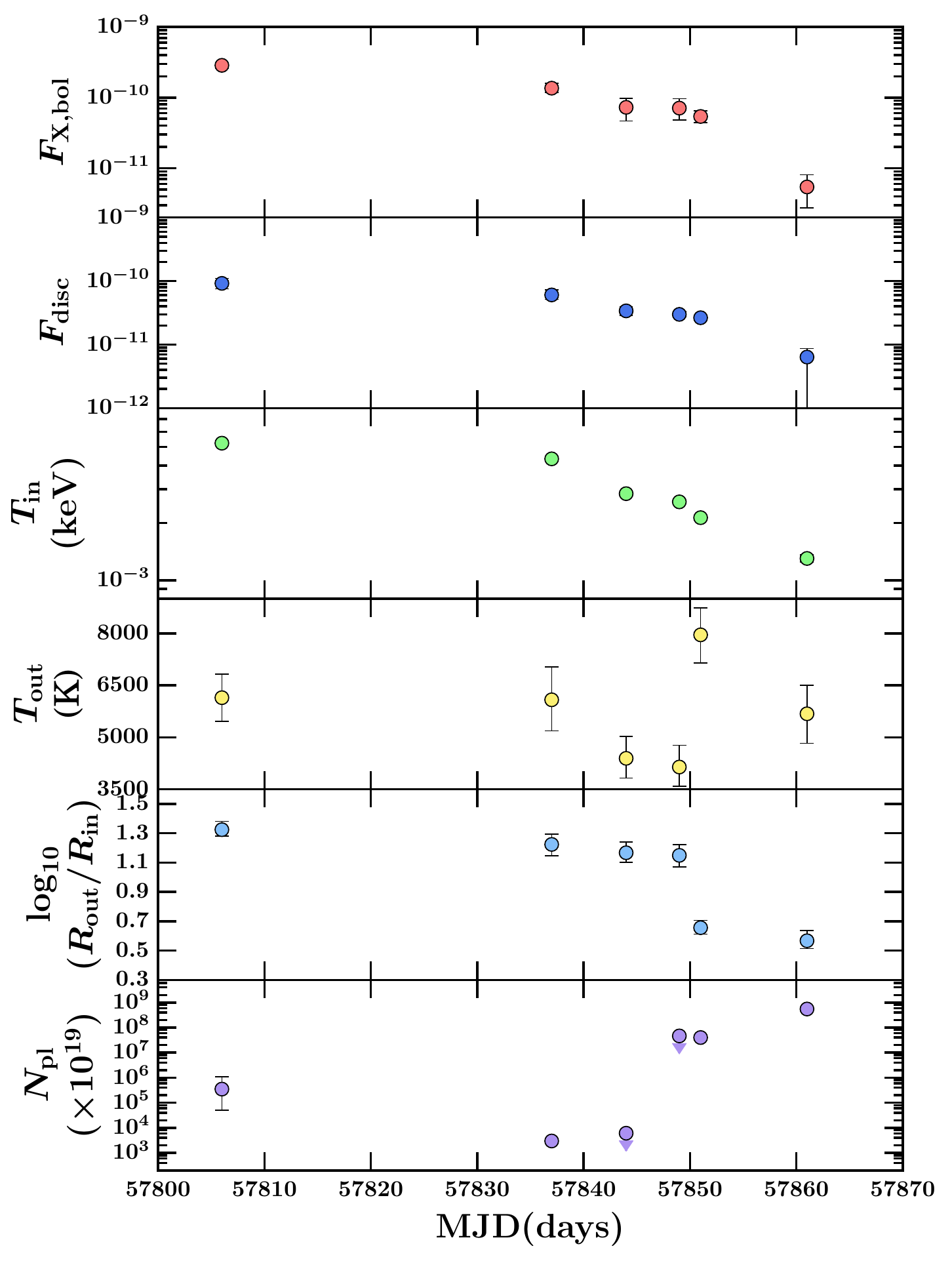}
\caption{Non-irradiated disc model fits to the J1753 mini-outburst. For each of our six epochs, we plot: bolometric X-ray flux (top panel), integrated disc flux (second panel), and the parameters derived from the MCMC algorithm fit with the non-irradiated disc model with temperature profile $T\propto R^{-3/4}$: inner disc temperature $T_{\rm in}$ (third panel), outer disc temperature $T_{\rm out}$ (fourth panel), log of the ratio of outer to inner disc radius $\log_{10}(R_{\rm out}/R_{\rm in})$ (fifth panel), and power-law normalization $N_{\rm pl}$ (bottom panel). The errors on bolometric flux include (90\% confidence) statistical instrument uncertainty only. 
Errors on fit parameters are $1\sigma$ confidence.
Upper limits on $N_{\rm pl}$ are displayed as purple arrows.
}
\label{fig:fit_params2}
\end{figure}

From our SED fitting, we are able to derive the time-series evolution of $T_{\rm in}$, $T_{\rm out}$, and $N_{\rm disc}$ over the course of the mini-outburst, for the irradiated disc model with two different temperature distributions (Fig. \ref{fig:fit_params}), and the non-irradiated disc model (Fig. \ref{fig:fit_params2}). We note the resulting posterior distributions of each parameter, in all epochs, are close to Gaussian in shape.
We find our six SEDs are well fit with an inner disc temperature that varies between $T_{\rm in}\sim0.03-0.2$ keV and $T_{\rm in}\sim0.01-0.1$ keV, for the irradiated $n=1/2$ and $n=3/7$ models, respectively. This is consistent with the range of $T_{\rm in}\sim0.1-0.4$ keV, found in previous X-ray spectral studies during the main (12 yr long) outburst of J1753, when the source was in the hard spectral state \citep{miller2006,hiemstra2009,chiang2010,reynolds2010,cassatella2012,kolehmainen2014,tomsick2015}. However, we note that the $T_{\rm in}$ is not well constrained by our data in the irradiated models (see Section \ref{sec:regimec}). 
For the non-irradiated ($n=3/4$) case, we find the six SEDs are fit with an inner disc temperature that varies between $T_{\rm in}\sim0.002-0.005$ keV ($\sim 2.3-5.8\times10^{4}$ K), significantly smaller than the irradiated cases.  

For the irradiated cases, our SED fits show the temperature in the outer disc decreases from $T_{\rm out}\sim1.1\times10^4-6.2\times10^3$K and $T_{\rm out}\sim1.2\times10^4-6.2\times10^3$K for the $n=1/2$ and $n=3/7$ models, respectively, as the source evolves from outburst maximum towards quiescence, with the exception of Epoch 5. Interestingly, we observe an increase in $T_{\rm out}$ during Epoch 5 for both irradiated disc models, the same epoch that shows the largest excess emission in the UV bands.
The decrease in $T_{\rm out}$ as the X-ray flux decreases during outburst decay is consistent with the predictions of the DIM+irradiation \citep{dubus1999,dubus2001}. 

\subsection{The X-ray Lightcurve}\label{sec:xr_lc}

\begin{figure*}
  \center
\subfloat[Long-term Light Curve Spanning May 2005 to April 2017]{\includegraphics[width=2.1\columnwidth]{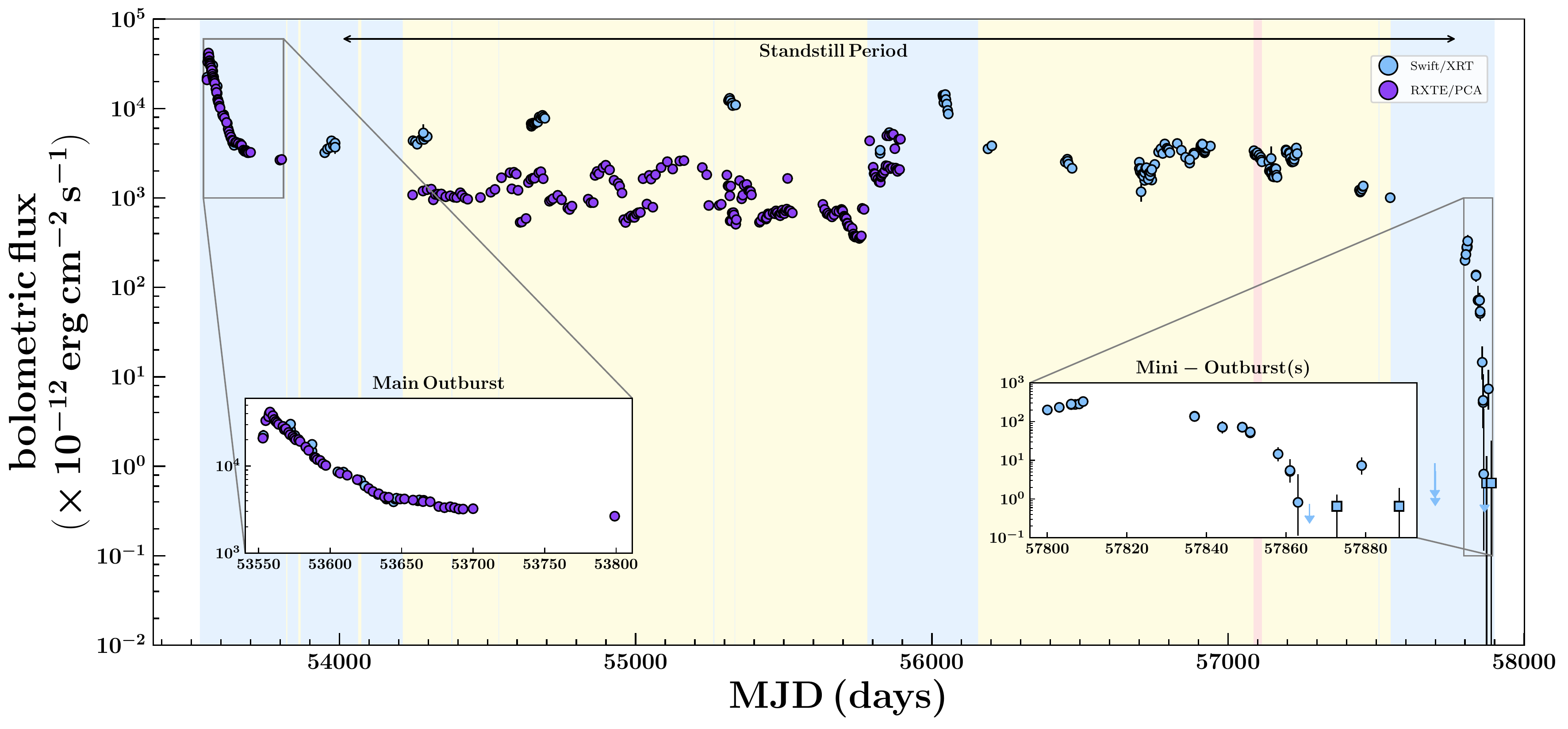}}\hfill
\subfloat[Main Outburst]
  {\includegraphics[width=.49\linewidth,height=.36\linewidth]{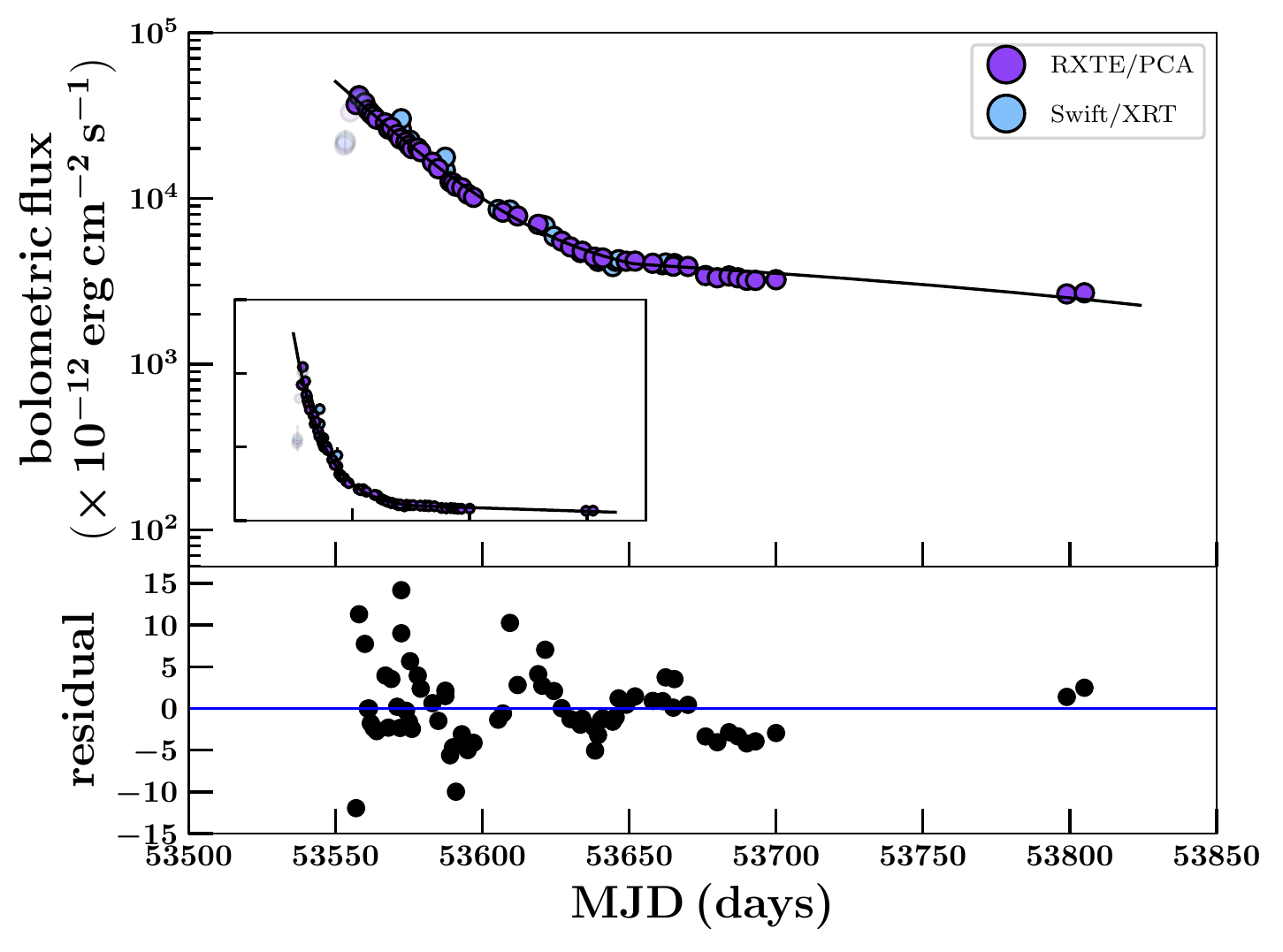}}\hfill
\subfloat[Mini-Outburst(s)]
  {\includegraphics[width=.49\linewidth,height=.36\linewidth]{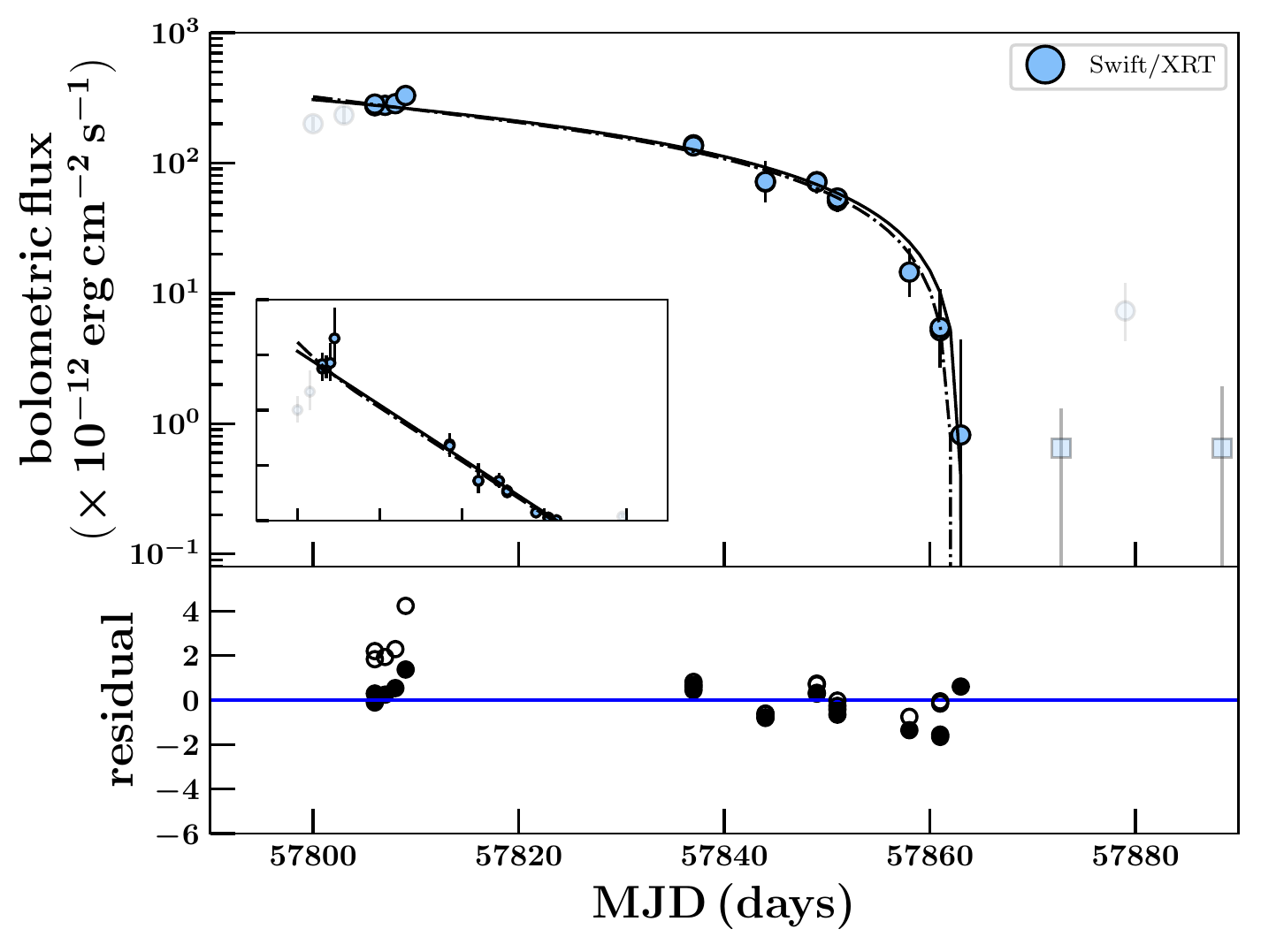}}\hfill
\caption{ 
Bolometric X-ray light curves of J1753: long-term behaviour (top), main outburst (bottom left) and mini-outbursts (bottom right). Data from individual instruments are colour coded (see legends). Error bars are individual (90\% confidence) instrument uncertainties. Upper limits are displayed with coloured arrows and low count rate flux estimates are represented by square markers (see Section \ref{sec:swift_data} for details). Background shading colours show the accretion state of J1753 as determined by \citet{tetarenko2016} and \citet{Shaw-2016b}: hard (blue), soft (red), and intermediate (yellow). See Section \ref{sec:longterm_data} for details.
Labelled on the long-term light curve (top) are  three accretion regimes: 
the main outburst, standstill period, and mini-outburst interval.
The black lines and lower panels (bottom figures) represent the best fit analytical model and residuals, respectively.
For the mini-outbursts (bottom right), we show the best fit pure linear (solid black, filled circle residuals) and exponential+linear (dot-dashed line and open circle residuals) decay models.
Translucent markers on these figures indicate the rise of the outburst, and second mini-outburst, not included in the fits. The inset axes display: zoomed in versions of the main outburst and mini-outburst regimes (top) and the lightcurves on a linear scale (bottom).
}
\label{fig:xr_lightcurve}
\end{figure*}

The X-ray light curves of recurring transient outbursts in LMXBs can be used as a powerful diagnostic to probe the physical mechanisms driving mass inflow/outflow in these systems \citep{tetarenko2018}. To this end, we have applied the Bayesian methodology developed by \citet{tetarenko2018} to the X-ray light curve of the J1753 mini-outburst, fitting their analytical irradiated disc-instability model to our data (see Fig. \ref{fig:xr_lightcurve}).
The DIM+irradiation predicts a multi-stage decay profile, starting with an exponential-shaped portion attributed to a viscously-accreting fully-irradiated disc, followed by a linear-shaped portion, occuring as a result of a cooling front propagating inward through the disc, at a rate set by the amount of irradiation heating (see \citealt{kingrit8,dubus2001}).

We find that the J1753 mini-outburst can be well fit by a pure linear-shaped decay on a timescale of $\tau_l=71\pm7$ days, and does not require the exponential+linear decay profile as in many Galactic BH-LMXB outbursts \citep{tetarenko2018b}. 
A pure linear profile allows us to calculate an upper limit on $C_{\rm irr}$ by assuming $f_t$, the flux level at which the transition occurs between the viscous (exponential-shaped) and irradiation-controlled (linear-shaped) decay stages in the light curve, is the maximum observed flux \citep{tetarenko2018b}. 
We find $C_{\rm irr}<4.2\times 10^{-2}$ (for an assumed $D=7.74_{-2.13}^{+5.00}$kpc).

Since an exponential decay stage could have occurred in the gap in our data around MJD $\sim57810 - 57838$, we also tested an exponential+linear decay profile. 
The best fit gives 
an exponential (viscous) 
timescale of $\tau_e=52_{-9}^{+8}$ days, a linear decay 
timescale of $\tau_l=48\pm9$ days, and a transition (between exponential and linear stages) occurring at time $t_{\rm break}=57814\pm9$ 
and flux level $f_t=(2.3\pm0.5)\times10^{-10} \, {\rm erg \, cm^{-2} \, s^{-1}}$. 
This exponential+linear profile allows us to calculate a $C_{\rm irr}$ from $f_t$, the flux level at which the transition occurs between the viscous (exponential) and irradiation-controlled (linear) decay stages in the light curve, of $C_{\rm irr}=(2.2_{-1.5}^{+3.3})\times 10^{-2}$ (for an assumed $D=7.74_{-2.13}^{+5.00}$ kpc).

\begin{table}
{
\setlength{\tabcolsep}{5pt}
\centering
\caption{ MCMC UVOIR/X-ray Correlation Fitting Results}
\medskip
\label{tab:c_fits}
\begin{tabular}{lccc} 
\hline
Wave &Central $\lambda$ & $m_{\nu}$ & $b_{\nu}$  \\
Band&($\rm{\mu m}$)&\\[0.045cm]
  \hline
  
$UVW2$&0.1928&$0.52_{-0.10}^{+0.14}$&$-0.25_{-0.56}^{+0.25}$\\[0.045cm]
$UVM2$&0.2246&$0.29_{-0.07}^{+0.08}$&$-0.59_{-0.16}^{+0.12}$\\[0.045cm]
$UVW1$&0.2600&$0.31_{-0.06}^{+0.05}$&$-0.68_{-0.14}^{+0.11}$\\[0.045cm]
$U$&0.3465&$0.25_{-0.04}^{+0.03}$&$-0.35_{-0.04}^{+0.03}$\\[0.045cm]
$B$&0.4392&$0.20_{-0.03}^{+0.04}$&$-0.18_{-0.03}^{+0.04}$\\[0.045cm]
$V$&0.5468&$0.22_{-0.04}^{+0.03}$&$-0.10_{-0.04}^{+0.13}$\\[0.045cm]
$R$&0.641&$0.08_{-0.08}^{+0.07}$&$-0.24_{-0.15}^{+0.17}$\\[0.045cm]
$I$&0.798&$0.08_{-0.07}^{+0.08}$&$-0.29_{-0.17}^{+0.18}$\\[0.045cm]
$J$&1.22&$0.05_{-0.06}^{+0.05}$&$-0.45_{-0.12}^{+0.15}$\\[0.045cm]
$H$&1.63&$0.04_{-0.08}^{+0.07}$&$-0.48_{-0.16}^{+0.20}$\\[0.045cm]
  \hline
 \multicolumn{4}{p{0.65\columnwidth}}{\hangindent=1ex NOTE. -- A linear fit was performed in log-space, where $m_{\nu}$ and $b_{\nu}$ are the slope and intercept, respectively.}\\
\end{tabular}\\
}
\end{table}

\subsection{UVOIR--X-ray Correlations}\label{sec:corr_X}

\begin{figure}
  \center
\includegraphics[width=\columnwidth,height=0.8\linewidth]{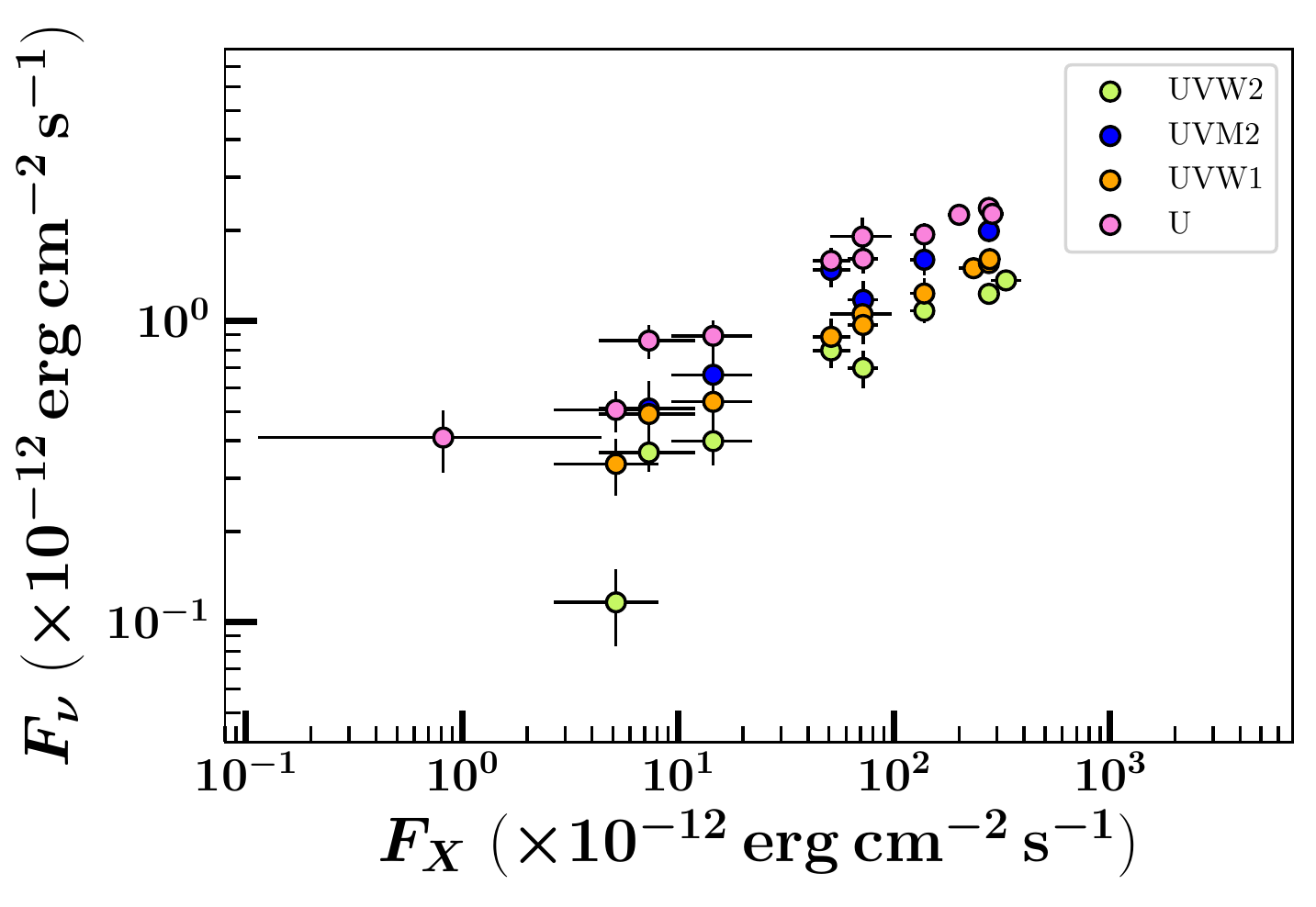}
\caption{UV--X-ray correlation during the J1753 mini-outburst. Dereddened flux in the available UV bands from {\em Swift}/UVOT is plotted vs. unabsorbed $2-10$ keV flux from {\em Swift}/XRT. Individual bands are colour-coded (see legend). Error bars include instrumental uncertainty only. 
}
\label{fig:UVX_corr}
\end{figure}

\begin{figure}
  \center
\includegraphics[width=\columnwidth,height=0.8\linewidth]{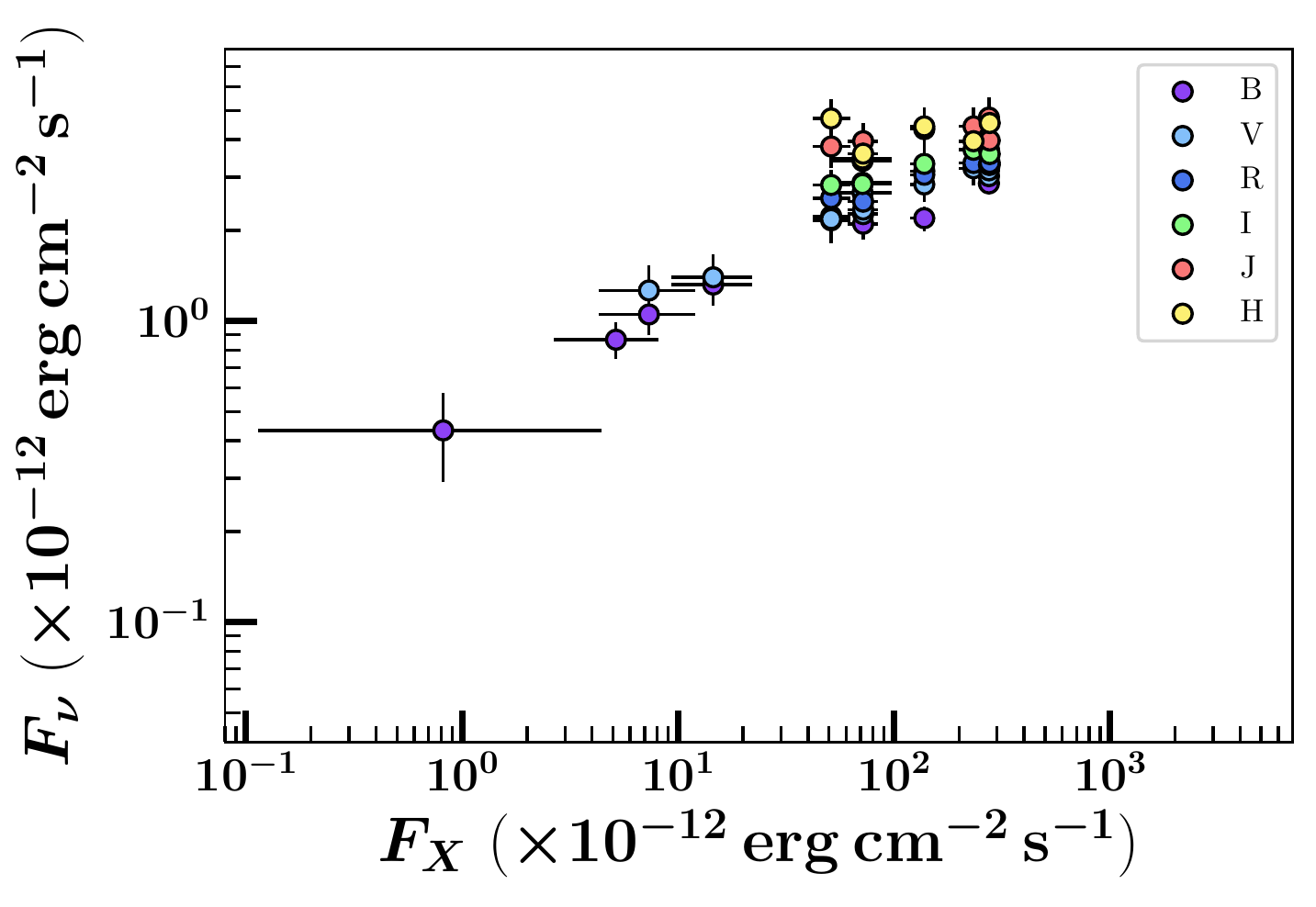}
\caption{Optical/IR--X-ray correlation during the J1753 mini-outburst. Dereddened flux in the available OIR bands from {\em Swift}/UVOT and {\em SMARTS} is plotted vs. unabsorbed $2-10$ keV flux from {\em Swift}/XRT. Individual bands are colour-coded (see legend). Error bars include instrumental uncertainty only. 
}
\label{fig:optX_corr}
\end{figure}

In addition to light curves and spectra, another way to quantify the contributions of different emission processes during an LMXB outburst is through the power-law correlations that exist between the flux at UVOIR wavelengths and the X-ray flux (\citealt{russell2006} and references therein). In Figs. \ref{fig:UVX_corr} and \ref{fig:optX_corr}, we plot $F_{\rm UV}$ and $F_{\rm opt/IR}$ versus $F_{X}$ for all simultaneous UV through X-ray data taken during the J1753 mini-outburst.
We  fit the correlations between the 10 individual UVOIR bands  during the mini-outburst and the $2-10$ keV X-ray flux (see Table \ref{tab:c_fits}) to determine the dominant emission processes in each waveband. 

We performed a linear fit in log-space to each dataset
with our Bayesian MCMC algorithm (see Section \ref{sec:mcmc_algorithm}),
and estimated the slope ($m_{\nu}$) and intercept ($b_{\nu}$) for each individual correlation.  
As the standard linear formulation (i.e., $y_{\nu} = m_{\nu}x + b_{\nu}$) is ill-suited for problems involving two-dimensional uncertainties, we use an alternative method, which parametrizes the slope 
in terms of the parameter $\theta$, defined as the angle that the linear function makes with the x-axis \citep{hog2010}.
After likelihood maximization is performed for $\theta$ and the y-intercept $y_b$, PDFs of $m_{\nu}$ and $b_{\nu}$ are obtained by taking the tangent of the resulting PDFs for $\theta$ and $y_b$.

We find that as the selected wavelength decreases, the slope of the correlation, $m_{\nu}$, increases. This trend has been found in outbursts of other short-period LMXBs containing both BHs and neutron stars  (e.g., Swift\,J1357.2-0933, SAX\,J1808.4-3658; \citealt{armaspadilla2013}; Beri, A. et al. 2018, submitted; \citealt{patruno2016}). This trend is expected for thermal emission. 

\section{Discussion - The Nature of the Long-term Behaviour in J1753}\label{sec:discuss}

The standard DIM+irradiation model cannot adequately describe the abnormal behaviour, characterized by outbursts of varying amplitude and duration, observed in the short-period BH-LMXB J1753. The division between transient and persistently accreting sources, predicted to occur by the DIM+irradiation in the $\Dot{M}-P_{\rm orb}$ plane, is based on whether an accretion disc is in the stable or unstable regime \citep{dubus1999}. An annulus $R$ within the disc can remain in the hot, stable state if the local accretion rate $\Dot{M}(R)$ is greater than the critical accretion rate of the hot state $\Dot{M}_{\rm crit}(R)$, which increases with radius. Thus, for the entire disc to remain in a stable hot equilibrium (i.e., present as a persistent source), the mass-transfer rate from the companion star ($\Dot{M_2}$) must be larger than the critical mass-transfer rate in the outer disc, $\Dot{M}_{\rm crit}(R_{\rm out})$. If $\Dot{M}_2<\Dot{M}_{\rm crit}(R_{\rm out})$, the disc will be in an unstable regime, undergo outbursts, and thus, be transient.

The critical accretion rate for an irradiated disc is parameterized as
\begin{align}
\nonumber \Dot{M}_{\rm crit}=9.5\times10^{14}C_{\rm irr,3}^{-0.36} \alpha_{0.1}^{0.04+0.01\log C_{\rm irr,3}} R_{\rm disc,10}^{2.39-0.10\log C_{\rm irr,3}} \cdot{}\\ M_1^{-0.64+0.08\log C_{\rm irr,3}}\, \rm{g/s},
\end{align}
\citep{lasota2008} where $M_1$ is the compact object mass in $M_{\odot}$, $R_{\rm disc, 10}$ is the disc radius in units of $10^{10}$ cm, $\alpha_{0.1}=\alpha/0.1$ is the $\alpha$-viscosity (a term parametrizing the efficiency of angular-momentum transport in the disc; \citealt{ss73}), and $C_{\rm irr,3}=C_{\rm irr}/10^{-3}$ is the irradiation constant (a parameter used to describe the fraction of the central X-ray luminosity that is intercepted and reprocessed by the disc; \citealt{dubus1999,dubus2001}).

\begin{figure*}
  \center
\includegraphics[width=1.6\columnwidth,height=0.9\columnwidth]{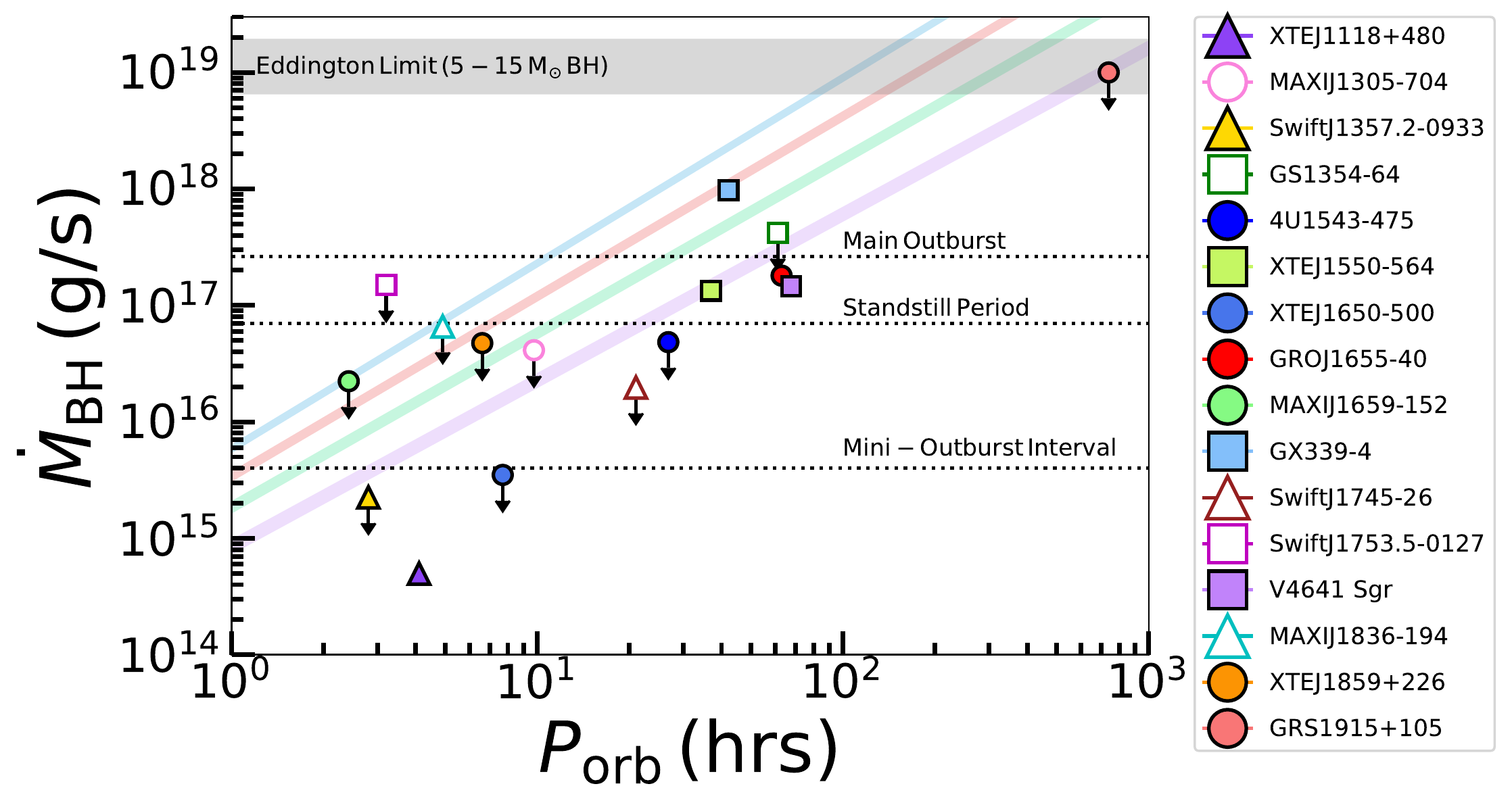}
\caption{Long term mass-transfer rate ($\Dot{M}$) vs.\ orbital period $(P_{\rm orb}$) for the transient Galactic BH-LMXB source sample, computed in the time-period 1996 January 6 -- 2015 May 14, from the WATCHDOG project \citep{tetarenko2016}. Mass-transfer rates  calculated assuming an  accretion efficiency $\eta=0.1$. 1$\sigma $ error bars are too small to see. Colours represent individual sources (see legend). Shapes 
denote accretion state(s) reached during outburst as determined by the WATCHDOG project: exclusively hard state or incomplete state transitions (source only reaches as far as the intermediate states) observed during all outbursts of the source (triangles), state transitions observed during all outbursts of the source (circles), or a combination of these two options (squares). 
Filled and open shapes represent sources with and without reliable distance estimates, respectively.
The mass-transfer rate estimates of sources that have only one detected outburst in the WATCHDOG time period are denoted as upper limits. The shaded grey region shows the Eddington limit for a $5-15$ $M_{\odot}$ BH. The shaded blue ($C_{\rm irr}=10^{-3}$), red ($C_{\rm irr}=5\times10^{-3}$), green ($C_{\rm irr}=3\times10^{-2}$) and purple ($C_{\rm irr}=3\times10^{-1}$) regions plot the critical accretion rate for an irradiated disc around a 5--$15 \, M_{\odot}$ BH according to the DIM+irradiation for various strengths of irradiation heating $C_{\rm irr}$. The dotted black lines show the average mass-transfer rate of J1753 calculated during each of the three accretion regimes. 
See Section \ref{sec:discuss} for details.}
\label{fig:mdot_porb}
\end{figure*}

In Fig. \ref{fig:mdot_porb}, we plot the long-term mass-transfer rate ($\Dot{M}_{\rm BH}$) as a function of  $P_{\rm orb}$ for the Galactic BH-LMXB source sample, computed in the time-period 1996 January 6 -- 2015 May 14, from the WATCHDOG project \citep{tetarenko2016}. Overlaid on the $\Dot{M}-P_{\rm orb}$ plane, we plot the (i) critical accretion rate for an irradiated disc, parametrized with four different choices of $C_{\rm irr}$, around a $5-15 M_{\odot}$ BH, and (ii) the average mass-transfer rate of J1753 during each of the three accretion regimes it traversed between May 2005 and April 2017 (dotted horizontal lines); see below for details).

The $\Dot{M}_{\rm BH}$ of J1753, averaged over a $\sim20$ year period, places it in the stable (persistent) region of the diagram. The nature of J1753's position in the plot has been discussed by multiple authors (see e.g., \citealt{coriat2012,tetarenko2016}).
As suggested by \cite{coriat2012}, for such a small disc ($P_{\rm orb}=3.2$ hrs or $R_{\rm disc}\sim7.4\times10^{10}$ cm) one of the most feasible ways to explain how J1753 stayed bright for so long, and in turn the sheer amount of material accreted during such an outburst is via a variable $\dot{M}_2$. 
An increase in $\dot{M}_2$ from its average value can keep the disc in a hot, stable state. We note that, there have been alternative explanations put forth to explain J1753, using the addition of tidal effects to the DIM (the tidal instability; \citealt{zur8,maccarone2013}), similar to what has been used to explain SU UMa stars \citep{osaki2013}. However, we do not focus on them in this paper.

This behaviour is reminiscent of what is observed in Z Cam stars \citep{simonsen2011,szkody2013}\footnote{Neutron star LMXBs have also been shown to exhibit Z Cam type behaviour. See \citealt{haswell2001}.}.  A subclass of dwarf novae, Z Cam stars are characterized by what is referred to as the ``standstill'' phenomenon (see e.g., \citealt{buatMenard2001}). In these systems, the decay from outburst maximum to quiescence is interrupted. Following the interruption, the luminosity of the system 
increases and then settles at an intermediate level, corresponding to some fraction of the original peak outburst luminosity.

This standstill, which can last anywhere from days to years at a time, ends when the system finally declines to the typical quiescent state once again. In the framework of the DIM, the standstill behaviour was originally interpreted as a stable phase of accretion \citep{osaki1974}, as a result of a mass-transfer rate from the companion star that fluctuates (either intrinsically or due to irradiation) about the critical mass-transfer rate, above which the disc is stable \citep{meyer1983,lin1985}. This conclusion was supported by various authors (see \citealt{buatMenard2001,hameury2014}), who were able to successfully model and reproduce the observed light curves of Z Cam systems by including a variable mass-transfer rate in their numerical DIM codes.

We test the hypothesis that J1753 is the BH-LMXB analog to Z Cam systems using the (i) available long-term X-ray light curves from {\it RXTE}/PCA and {\it Swift}/XRT. (see Fig. \ref{fig:xr_lightcurve} and Section \ref{sec:longterm_data}) and (ii) (quasi) simultaneous UVOIR and X-ray data from discrete intervals, during the time period of May 2005 to April 2017. To do so, we split the $\sim$ 12 years of outburst activity into three accretion regimes: the main outburst (A), standstill period (B), and mini-outburst interval (C). 

\subsection{Regime A: The Main Outburst}

Regime A can be attributed to a typical transient outburst, predicted by the irradiated DIM, in which the accumulation of matter in the disc triggers the thermal-viscous instability and subsequent cycling of the disc into a hot, ionised outburst state over a viscous timescale.


\subsubsection{X-ray Light Curve Fitting}
As the disc dissipates through viscous accretion of matter at this point, we expect to see an exponential-shaped decay profile in the X-ray light curve. 
As described in Section \ref{sec:xr_lc} for the mini-outbursts, we model the observed bolometric X-ray light curve of the main outburst using the analytical version of the DIM+irradiation \citep{dubus2001} built by \cite{tetarenko2018}. 
The light curve (Fig. \ref{fig:xr_lightcurve}) is well fit with an exponential+linear shaped decay profile.
The exponential 
timescale is $\tau_e=117\pm1$ days\footnote{Using the methodology developed by \citealt{tetarenko2018} and this viscous timescale, we find the light curve profile corresponds to an $\alpha$-viscosity parameter of $\alpha=0.16\pm0.03$. Interestingly, this is consistent with the $\alpha\sim0.1-0.2$ inferred in dwarf novae \citep{kotko2012}.}, and the estimated mass-transfer rate from the companion is  $\dot{M}_2=(2.63_{-0.04}^{+0.04})\times 10^{17}$ g/s, 
significantly higher than the critical mass-transfer rate in the outer disc of J1753, $\dot{M}_{\rm crit}=(1.6_{-0.2}^{+0.4})\times10^{16}$ g/s (calculated for a standard $C_{\rm irr}=5\times10^{-3}$). 

The standard DIM+irradiation interpretation \citep{dubus2001} attributes the linear-shaped (irradiation-controlled) portion of the decay profile to a cooling front propagating inward through the disc, where the speed at which this front propagates is controlled by the decaying X-ray irradiating flux. In this interpretation, the transition from the exponential (viscous) to the linear (irradiation-controlled) stage of the decay occurs when the irradiation temperature at the outer radius drops below $T_{\rm out}\approx 10^4$ K, the temperature at which hydrogen starts to recombine. \cite{tetarenko2018} find the flux level of this transition in short-period ($P_{\rm orb}<5$ hr) BH-LMXB systems to vary between $f_t\sim1\times10^{-10}-3\times10^{-9} \, {\rm erg \, cm^{-2} \, s^{-1}}$. From our light curve modelling, we find (i) this transition to occur at time $t_{\rm break}=536512_{-1}^{+2}$ and flux level $f_t=(4.03\pm0.04)\times10^{-9} \, {\rm erg \, cm^{-2} \, s^{-1}}$, and (ii) a linear decay timescale of $394_{-17}^{+18}$ days. 

Here, we postulate a scenario, similar to what is described in \cite{hameury2014} for Z Cam stars, whereby the $\dot{M}_2$ decreases, allowing  $T_{\rm out}$ to drop below $10^4$ K, triggering the development of an inward propagating cooling front in the outer disc. If $\dot{M}_2$ remained low, and the inward movement of the cooling front (and thus observed X-ray decline) was purely controlled by the decaying X-ray irradiating flux, we would expect the source to drop into quiescence at $\sim54046$. However, J1753 did not follow this predicted pattern.
 
\subsection{Regime B: The Standstill Period}\label{sec:regimeb}

After several months spent in a linear decay stage (at near constant flux level), the flux began gradually increasing, until MJD$\sim54500$, where a sudden increase in flux was observed (see \citealt{Soleri-2013,Shaw-2013a} for details). 
J1753 then spent $\sim9$ years at an intermediate flux level, before dropping into quiescence at MJD$\sim57698$ \citep{Shaw-2016c}. 
We 
connect 
this standstill period, regime B, to the standstill phenomenon observed in Z Cam stars (see \citealt{hameury2014}). We postulate that, at some point, $\dot{M}_2$ 
increased, before the cooling front had a chance to reach the inner edge of the disc. At this point, an outburst beginning in the outer disc (an ``outside-in'' outburst) began, as a result of the enhanced mass transfer, before the disc had a chance to reach quiescence.

\subsubsection{Quantifying Mass Transfer}

To determine if this scenario is plausible, we have first computed an average mass-transfer rate during the standstill period (which we approximate as $54000-57698$). Using the algorithm presented in \cite{tetarenko2016}, we estimate an average mass-transfer rate by computing the time-averaged bolometric luminosity using the 
X-ray light curve data
for this time period
available from RXTE/PCA and Swift/XRT (see Fig. \ref{fig:xr_lightcurve} and Section \ref{sec:longterm_data} for details),  assuming $D=7.74_{-2.13}^{+5.00}$ kpc and a fixed accretion efficiency $\eta=0.1$. We find
$\dot{M}_2=6.2_{-3.4}^{+11.6}\times 10^{16}$ g/s,
which is greater than $\dot{M}_{\rm crit}$ in the outer disc, supporting the standstill phenomenon scenario.

\subsubsection{UVOIR SED Fitting}\label{sec:regimeB_SED}

\begin{figure*}
  \center
\includegraphics[width=.33\linewidth,height=.23\linewidth]{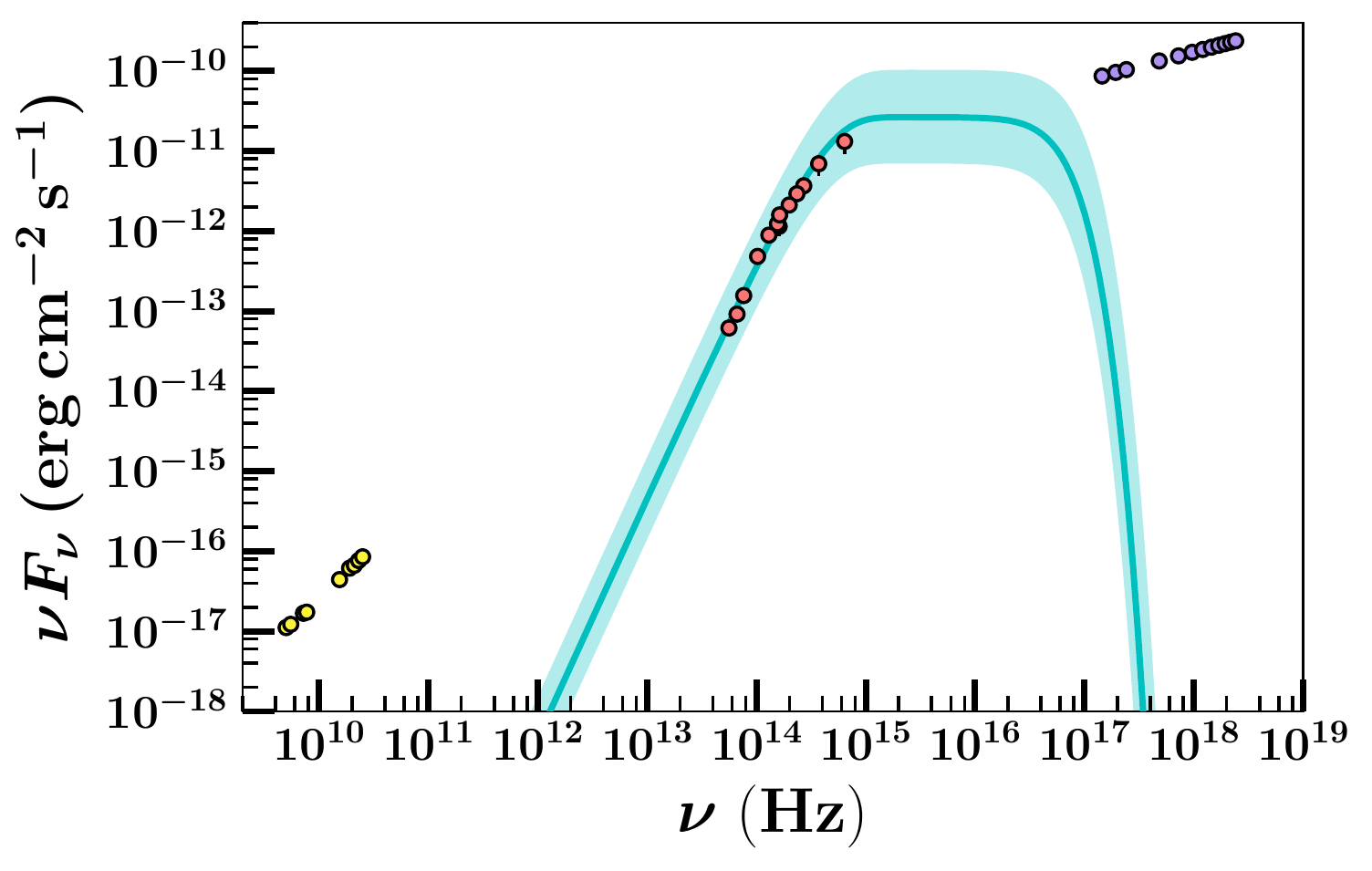}\hfill
\includegraphics[width=.33\linewidth,height=.23\linewidth]{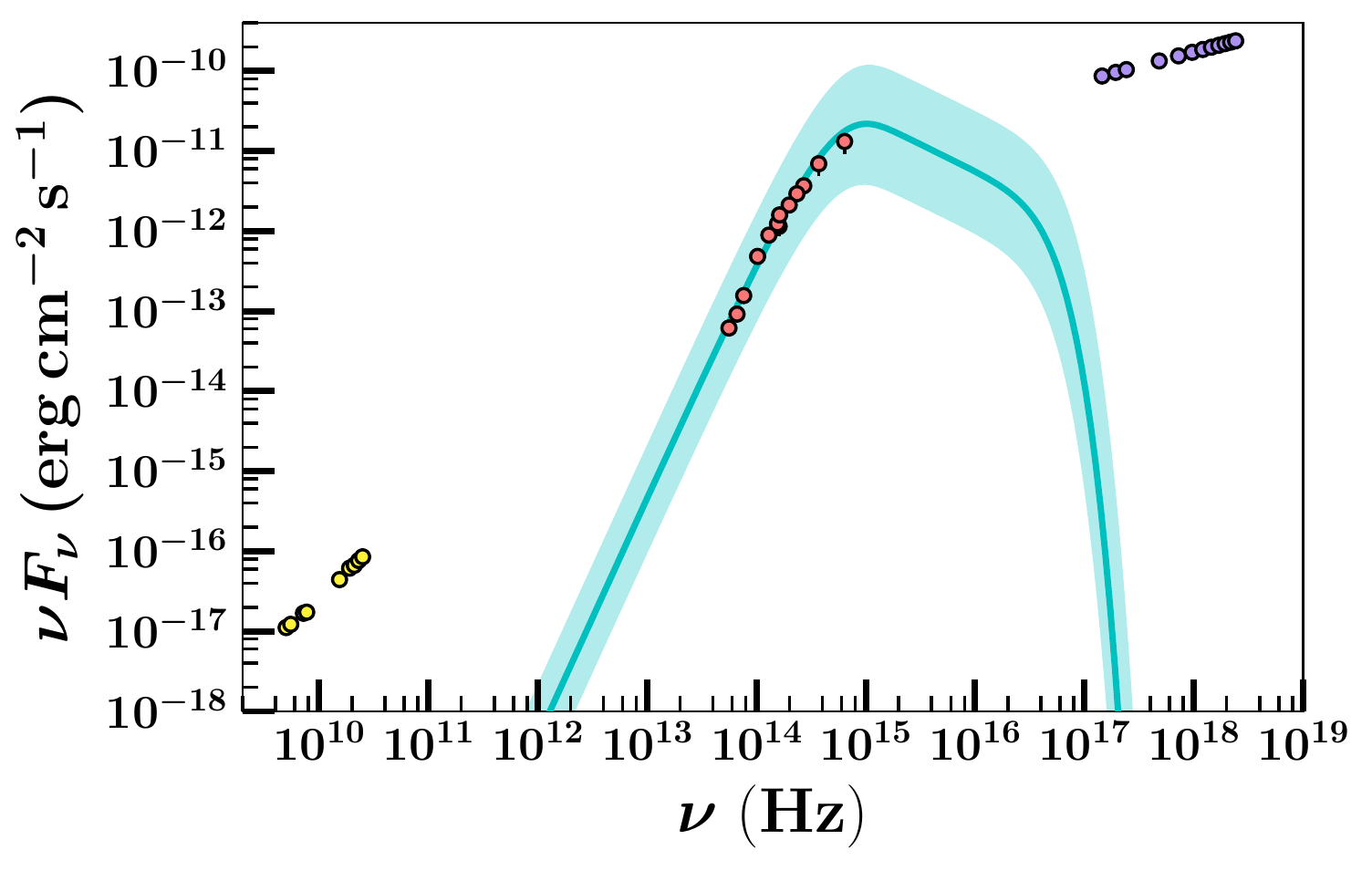}\hfill
\includegraphics[width=.33\linewidth,height=.23\linewidth]{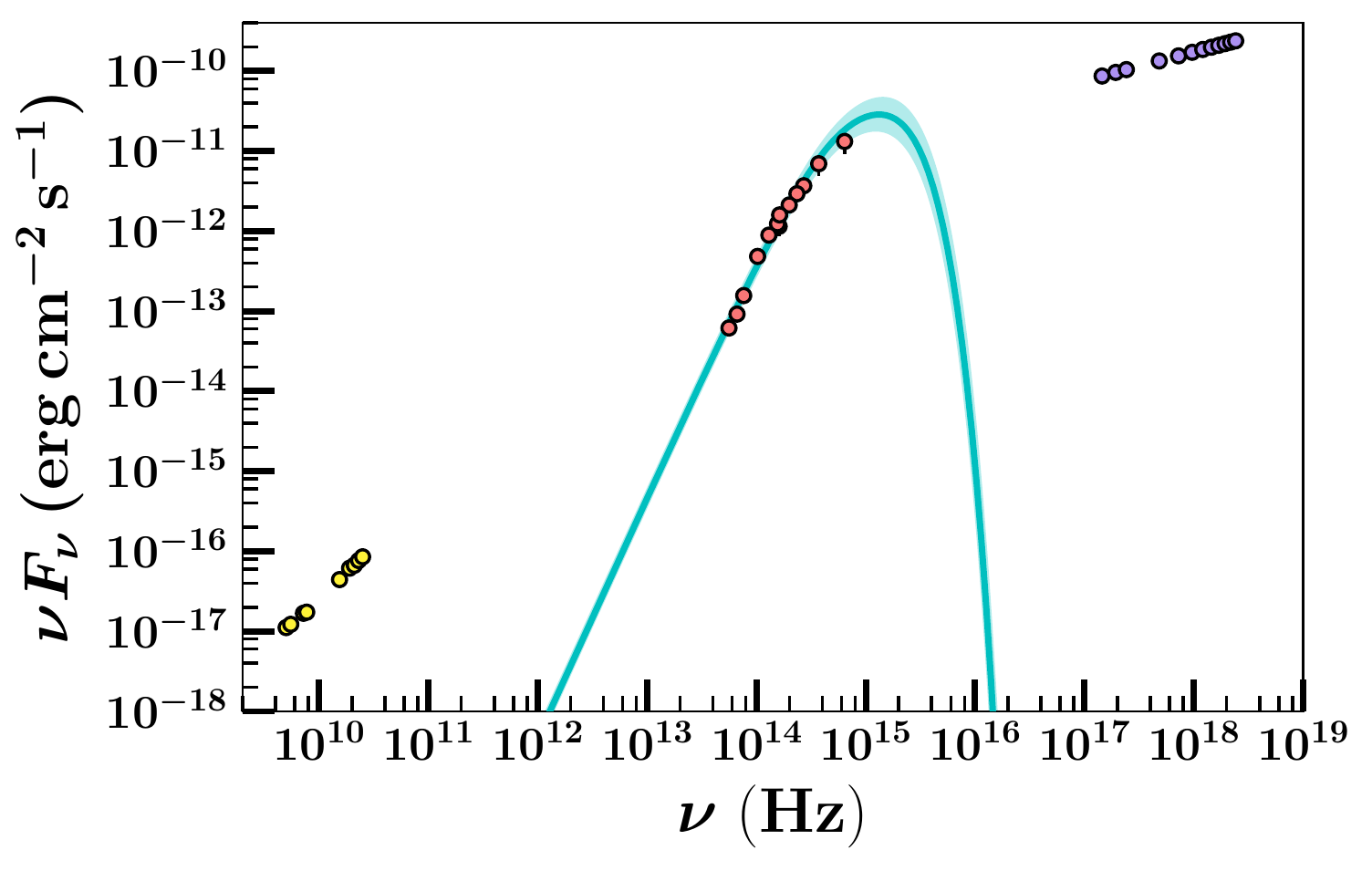}\hfill
\caption{Broadband SED, during one epoch of the standstill period of SwiftJ1753.5-0127 (April 4-5, 2014; 56751-56752),
fit with the irradiated disc model with a temperature distribution $T\propto R^{-n}$ characterized by $n=1/2$ (left), $n=3/7$ (middle), and a non-irradiated disc model with a temperature distribution characterized by $n=3/4$ (right). All data are simultaneous or quasi-simultaneous (within 1 day)
with the exception of K-band which was taken two days prior. All data have been dereddened. The solid cyan line and shaded regions represent the best-fit and 1$\sigma$ confidence intervals from the MCMC fitting algorithm, respectively.
Only the UVOIR data (red) are fit. X-ray ({\em Swift}/XRT; purple) and radio (VLA and AMI; yellow) are also plotted to show the multi-wavelength behaviour of the source.}
\label{fig:2014_SED}
\end{figure*}

In addition, we have also fit the UVOIR SED, made up of simultaneous data obtained during one epoch of this standstill period (April 2014; \citealt{tomsick2015}), with our disc models. See Table \ref{tab:outfits} for best-fit results and Fig. \ref{fig:2014_SED}. 

Overall, we find that the 2014 SED is consistent with a cool disc, as reported by \citet{tomsick2015}, with no evidence of a jet in the NIR.
Despite the much higher bolometric flux in comparison to the mini-outburst period, $T_{\rm out}$, in the irradiated cases, is consistent with values during the decay of the first mini-outburst (see Section \ref{sec:regimec} below). In addition, we find a similar pattern for $T_{\rm in}$ and $N_{\rm disc}$ in the 2014 observations, compared to the mini-outbursts, in both irradiated cases. 


From the SED fit, we also derive an integrated disc flux of $F_{\rm disc}=(0.3-4.6)\times 10^{-10} {\rm erg \, cm^{-2} \, s^{-1} }$ and $F_{\rm disc}=(0.1-3.3)\times 10^{-10} {\rm erg \, cm^{-2} \, s^{-1} }$ for the irradiated $n=1/2$ and $n=3/7$ models, respectively. Combining knowledge of $F_{\rm disc}$ with the simultaneous X-ray data allows for an estimate of the irradiation constant $C_{\rm irr}$, defined as the fraction of the X-ray luminosity intercepted and reprocessed by the disc. We find $C_{\rm irr}\sim8.8\times10^{-2}$ and $C_{\rm irr}\sim4.6\times10^{-2}$ for the $n=1/2$ and $n=3/7$ models, respectively. Both estimates are significantly larger\footnote{It is worth noting that a large fraction of intercepted X-rays ($C_{\rm irr}$) has been found to be consistent with observations in multiple other BH-LMXBs as well \citep[e.g.][]{gandhi2010,tetarenko2018b}.} than the standard value typically assumed in LMXBs ($5\times 10^{-3}$; \citealt{dubus2001}), suggesting it is possible that strong X-ray irradiation of the companion star could be the cause of this period of enhanced $\dot{M}_2$ in the system.

The changes in observed spectral properties of J1753, from the main outburst regime through the standstill phase, are also consistent with the scenario implied by our UVOIR SED fitting, whereby the disc truncates and recedes from the BH as the outburst evolves.
\citet{kajava2016} find that the evolution of spectral changes, observed in J1753 from the main outburst through the standstill phase, are consistent with being driven by the truncation of the disc and subsequent formation of a large hot inner corona inside it, causing the dominant source of X-ray emission (i.e., seed photons for Comptonization) to switch from the disc (via up-scattering) to self-produced synchrotron emission (via syncrotron self-Compton) in the hot corona.

Finally, we note that the 2014 data can also be fit well with a non-irradiated disc model. In this case, the scenario would be a thin viscous disc, truncated at a high inner radius, with presumably a hot radiatively inefficient flow inside producing the hard X-ray emission. 
However, from the SED fit with the $n=3/4$ model, the derived integrated disc flux of $F_{\rm disc}=0.5_{-0.2}^{+0.4}\times 10^{-10} {\rm erg \, cm^{-2} \, s^{-1} }$, corresponds to a mass-accretion rate $>10^{19}$ g/s (for an assumed $D=8$ kpc and an inclination $i<80$ degrees), far too large to be consistent with the truncated thin disc scenario. Thus, we consider the non-irradiated disc model to be unphysical.

\subsection{Regime C: The Mini-Outburst Interval}\label{sec:regimec}

Finally, after a standstill period, J1753 returned to a quiescent state on 2016 November 6 (57698). However, it would not remain in quiescence for long. $\sim102$ days later ($\sim57800$), J1753 was observed to undergo two low-luminosity mini-outbursts before returning to a quiescent state for the second time. We attribute the mini-outburst interval, regime C, to $\dot{M}_2$ remaining low enough (i.e., below $\dot{M}_{\rm crit}$), putting the disc in the instability zone, and allowing for a series of transient outbursts to occur.

\subsubsection{Quantifying Mass Transfer}

To determine if this scenario is plausible, we have first computed an average mass-transfer rate (with the same method described in Section \ref{sec:regimeb} above), during the period ranging from when J1753 first returned to a quiescent state ($\sim57698$) until the end of the two mini-outbursts ($\sim57885$). We find 
$\dot{M}_2=3.6_{-3.3}^{+13.5}\times 10^{15}$ g/s,
which lies below $\dot{M}_{\rm crit}$ in the outer disc of J1753, with the upper error bar on par with $\dot{M}_{\rm crit}$.

\subsubsection{X-ray Light Curve Fitting}\label{sec:xrlc_regimeC}

As described in Section \ref{sec:xr_lc}, we model the observed bolometric X-ray light curve of the first-mini-outburst using the analytical version of the DIM+irradiation built by \cite{tetarenko2018}. 
As there is a gap in the data between MJD $\sim57810 - 57838$, during which time an exponential (viscous) decay may have taken place prior to the linear decay, we attempt to fit the data (see Fig. \ref{fig:xr_lightcurve}) with both a
pure linear-shaped decay profile and an exponential+linear shaped decay profile.

In the context of the DIM+irradiation, if the decay is truly only linear, this would imply that the heating front never reached the outer disc. In the standard DIM, an ``inside-out'' heating front can stall somewhat easily \citep{dubus2001}, leading to short, low-amplitude outbursts that take place in the innermost regions of the disc. This is because the critical density needed to raise a ring of accreting matter to the hot, ionized state increases with radius. So, if the outward moving heating front is unable to raise the density above this critical value, the outburst stalls and a cooling front will develop. In the DIM+irradiation, irradiation heating will reduce the critical density, allowing the heating front to reach larger radii. However, this assumes that the irradiating flux is seen by the entire disc, i.e. there is no self-screening of the central X-ray source by the inner disc \citep{dubus2001}. 

This assumption is unlikely to be true in J1753. The estimates of $C_{\rm irr}<4.2\times10^{-2}$ (pure linear decay) and $C_{\rm irr}=(2.2_{-1.5}^{+3.3})\times 10^{-2}$ (exponential+linear decay), from the X-ray light curves fitting, are a factor $\sim 5$ higher than the standard value of $C_{\rm irr,expected}\sim 5\times 10^{-3}$, yet still compatible with the stability limits between transient and persistent LMXBs \citep{coriat2012}. If the irradiation source in J1753 is large, causing X-rays to encroach on the disc vertically (e.g. via a hot, inner corona flow), this could result in an intercepted fraction that is high. This, combined with the low peak flux and short duration of the mini-outburst, leads us to favour the exponential+linear shaped decay in the X-ray light curve, which supports the scenario, implied by the UVOIR SED behaviour (see Section \ref{SED_discuss} below), of a fully-irradiated, truncated disc, heated by a source of irradiating X-rays produced in a corona above the disc.

\subsubsection{UVOIR SED Fitting}\label{SED_discuss}

We have also fit the UVOIR SEDs, made up of simultaneous/quasi-simultaneous data obtained during six individual epochs of the first mini-outburst. See Fig. \ref{fig:broadband_seds} and Table \ref{tab:outfits}. Figs \ref{fig:fit_params} and \ref{fig:fit_params2} show the evolution of $T_{\rm out}$ in the outer disc throughout the course of the mini-outburst for the irradiated and non-irradiated cases, respectively. 

In the irradiated cases, our UVOIR SED fits imply that, overall, the outer disc cools as the source heads toward quiescence (prior to the second mini-outburst). According to the predictions of the DIM+irradiation, when $T_{\rm out}<10^4$ K, the temperature at which hydrogen ionizes, a cooling front propagates through the disc, resulting in the source flux declining towards quiescence. This behaviour is echoed in Fig. \ref{fig:fit_params}. In conjunction with this cooling, we also note an upward trend in $N_{\rm disc}$ (and by extension $R_{\rm in}$; Table \ref{tab:outfits}),  indicating that the inner disc recedes from the BH as the mini-outburst progresses. The simultaneous decrease of $T_{\rm in}$ is consistent with this recession, indicating that the inner edge of the disc cools as it moves further away from the BH. 

Interestingly, we observe a strong increase in $N_{\rm disc}$ in the last two epochs, indicating that there must also be a strong increase in $R_{\rm in}$ (see Table \ref{tab:outfits}). Working under the assumption that $R_{\rm out}$ doesn't change much over the mini-outburst, this observation is compatible with the decrease observed in $\log_{10}(R_{\rm out}/R_{\rm in})$, and suggests that $R_{\rm in}$ increased by a factor 4-8 (depending upon the choice of $n$, see Table \ref{tab:outfits}) in $\sim2$ days. This large change seems to coincide with the drop in optical magnitude prior to the second mini outburst (see Fig. \ref{fig:stacked_lc}).

Although the overall trend is one of a cooling disc as J1753 heads toward quiescence, in epoch 5 there is an apparent increase in $T_{\rm out}$ with respect to epoch 4. In epoch 6, $T_{\rm out}$ returns to a level consistent with epoch 4, as the cooling continues. Interestingly, the SED of J1753 at epoch 5 (Fig. \ref{fig:broadband_seds}) shows evidence of a significant excess (with respect to the best-fit model) in the UV bands. In various dwarf novae, a similar UV excess is observed (see e.g., \citealt{smak1999,smak2000,hameury2000}). This excess is attributed to brightening of the hot spot, where the mass transfer stream interacts with the outer disc, as a result of mass transfer variations from the companion. This is unlikely to be the case in J1753, as the irradiated disc would likely dominate over hot spot emission in the UV bands.


From the SED fits, we  derive an integrated irradiated disc flux of $F_{\rm disc}\gtrsim F_{\rm X,bol}$ in all epochs, implying an unphysical $C_{\rm irr}>1$. While this 
suggests that the data are not consistent with an irradiated disc, it is more likely that we are just overestimating $F_{\rm disc}$, as we discuss below.

In the irradiated disc fits, the $T_{\rm in}$ parameter is not well constrained. The break between the flat power law and the Wien exponential cutoff in the spectrum is set by $T_{\rm in}(R_{\rm in})$. Looking at the fitted irradiated disc SEDs (see Fig. \ref{fig:broadband_seds}), it is clear that a disc truncated at a larger $R_{\rm in}$, which would effectively change the higher end of the spectrum without changing the lower end, could also fit the data. A smaller $T_{\rm in}$ parameter (and thus a disc truncated at a larger $R_{\rm in}$) would lead to a lower $F_{\rm disc}$. The (i) X-ray spectrum throughout the mini-outburst, fit well with a power law and no viscous disc-blackbody needed, and (ii) the large $C_{\rm irr}$ parameter estimated from the X-ray light curve fitting, implying an irradiation source in a hot, inner corona flow, both support this truncated irradiated disc scenario. 

The other possibility to explain the large $F_{\rm disc}$ estimates is that only a part of the disc was hot during the mini-outbursts. The SED model explicitly assumes that $T_{\rm out}$ is measured at the outer radius of the full disc, $R_{\rm out}$. If the heating front never reached the outer edge of the disc during the mini-outburst,
the temperature $T_{\rm out}$ we measure would refer to the maximum radius of the hot disc $R_{\rm hot}<R_{\rm out}$, and the true value of $F_{\rm disc}$ would be smaller. 
However, we note that this scenario is at odds with the large $C_{\rm irr}$ we estimate from the light curves.

Similar to the 2014 data, we note that the mini-outburst can also be fit well with a non-irradiated disc model. 
However, the derived integrated disc fluxes in all epochs correspond to mass-accretion rates $>5\times10^{18}$ g/s (for an assumed $D=8$ kpc and an inclination $i<80$ degrees). As discussed in \ref{sec:regimeB_SED} regarding the 2014 data, these mass-accretion rates are far too large to be consistent with the truncated thin disc scenario. Thus, we consider the non-irradiated disc model to not be a viable option to describe the mini-outburst data.

\subsubsection{UVOIR--X-ray correlation}\label{corr_x_regimeC}

In a study of 33 LMXBs, \citet{russell2006} found 
a global correlation with a slope $m_{\nu}=0.6\pm0.1$ for $F_{\rm opt/IR}\propto F_X^{m_{\nu}}$ for BH-LMXBs in a hard accretion state. This correlation, which 
extends to the UV regime \citep[e.g.][]{Rykoff-2007}, is instrumental in our understanding of the emission processes in the accretion disc. The value of $m_{\nu}$ can vary significantly, depending on the dominant emission mechanism. \citet{vanp1994} find that $m_{\nu}=0.5$ is expected for emission dominated by X-ray reprocessing in the disc, whereas $m_{\nu}=0.7$ implies the presence of an optically thick synchrotron jet \citep{russell2006}. For a viscously heated disc, we expect $m_{\nu}$ in the range $0.15 - 0.3$ in the NIR to UV bands \citep{russell2006}. 
 
Figs. \ref{fig:UVX_corr} and \ref{fig:optX_corr} show the UVOIR fluxes plotted against their corresponding X-ray flux (in the 2--10 keV band) during the first mini-outburst. The results of a linear fit (in log-space) to the data in each UVOIR band are presented in Table \ref{tab:c_fits}. As discussed in Section \ref{sec:corr_X} we find a decreasing slope as we move to redder filters. A dependency of $m_{\nu}$ on wavelength is predicted by theory \citep{Frank-2002}, and has been observed in the short period BH-LMXB Swift\,J1357.2-0933 \citep{armaspadilla2013}. However, unlike Swift\,J1357.2-0933, in which the slope was consistent with a viscous disc in all {\em Swift}/UVOT filters, we see an apparent switch from a viscous disc to reprocessed emission in the $UVW2$ band. 

The fact that we see from the UVOIR-X-ray correlations, that only the $UVW2$ flux is consistent with reprocessing, could suggest that the entire disc is not being irradiated and thus, there is a screening process preventing the irradiating flux from reaching the outer radii of the disc. In this case, irradiation alone would not bring the entire disc into a hot, ionized state. As a result, the heating front would never reach the outer edge of the disc during the mini-outburst, resulting in an outburst from only a portion of the disc.

However, this scenario is inconsistent with the large $C_{\rm irr}$ estimates from the X-ray light curve fits, which imply a large source of X-ray irradiation (possibly via the hot, inner corona flow). We also note that the best-fit value for $m_{\nu}$ in the $UVW2$ band appears to be driven by one data point at a low flux, highlighted by the relatively large uncertainties compared to those in the other filters. If the low-flux point is removed, then the $UVW2$-X-ray correlation would be consistent with the viscous disc scenario implied by other filters. We therefore are cautious to interpret the $UVW2$ emission as being a result of X-ray reprocessing alone.

The likely reality is that there are a number of emission processes contributing to the optical and UV flux of J1753 during the mini-outburst.
In addition to supporting the scenario where both the disc geometry and the source of X-ray emission change as J1753 evolves from the main outburst through the standstill phase, the evolution of the timing properties of the source have also been shown to be consistent with the existence of multiple sources of optical/UV emission. Recently, \citet{Veledina-2017}  attributed the dramatic changes  in the shape of the optical/X-ray cross-correlation functions (CCFs), as J1753 evolved from the main outburst to the standstill phase, as the source of optical/UV emission switching from purely reprocessed X-ray photons in the outer regions of the disc (during the initial stages of the main outburst), to a combination of emission from the irradiated disc and synchrotron emission from the hot inner corona (following the drop in X-ray flux and the subsequent change in the source of seed photons producing the comptonized X-ray emission in the standstill phase).

The shallower (than expected for emission from a pure irradiated disc) optical/UV-X-ray correlations we observe during the mini-outburst are consistent with this scenario, whereby the optical/UV emission is a combination of X-ray reprocessing in the outer disc and a (likely significant) contribution of (optical/UV) synchrotron emission from the
corona, effectively reducing the slope we observe and fit. 



In the NIR $J$ and $H$ bands, the slope is consistent with 0, implying that there is no NIR response to the changing X-ray flux. We find that a jet component is required to describe the NIR emission in four of the six SEDs (Fig. \ref{fig:broadband_seds}), indicating that the jet may be contaminating the NIR-X-ray correlations. This is similar to the BH-LMXB XTE\,J1550-564, which underwent a mini-outburst in 2003 and exhibited evidence of a compact jet in the NIR portion of the SED \citep{chaty2011}, and has also been seen in J1753 previously \citep{rahoui2015}.

The $\beta\sim1.2$ spectral index of the PL component found to fit the NIR emission in our SEDs is steeper than typically seen in optically thin radio jets \citep[$\beta\sim0.6$ e.g.][]{Russell-2010} but not unphysical (see e.g., discussion in \citealt{tetarenkoa2017}). We must note that the best-fit value of $\beta$ is only driven by one or two NIR data points, so is likely less constrained than the MCMC derived uncertainties suggest. Therefore we suggest that the contribution from the jet is likely contaminating the NIR-X-ray correlation. 



\section{Summary and Conclusions}
\label{sec:conclusions}
We have presented here a comprehensive study of the BH-LMXB Swift J1753.5-0127, focusing in particular on multi-wavelength data obtained during a mini-outburst toward the end of its $\sim12$ year long outburst. By modelling the observed UVOIR SED at six epochs during the mini-outburst with an irradiated accretion disc we were able to track the evolution of some important physical properties of the disc. Additionally, we modelled the profile of the X-ray light curve during the mini-outburst, enabling us to probe the mechanisms of mass transport in the disc as well as the X-ray irradiation properties. Finally, we utilised our wealth of (near-) simultaneous multi-wavelength data to investigate the correlation between the observed X-ray and UVOIR fluxes. These correlations allowed us to investigate the emission mechanisms present during the mini-outburst.

Ultimately, we conclude that the evolution of the mini-outburst is consistent with a scenario involving a fully irradiated disc, truncated at a large radius, with a hot, radiatively inefficient, accretion flow inside it, within which the source of irradiating X-rays (likely produced mainly by the synchrotron self-Compton process), and a significant portion of the optical/UV (via synchrotron photons) emission, are produced.

This scenario is supported by the observed (i) light curve profile, displaying the classic exponential+linear shape predicted by the DIM+irradiation \citep{dubus1999,dubus2001} for the outburst of an irradiated disc, (ii) high fraction of reprocessed X-rays ($C_{\rm irr}$) derived from the X-ray light curve, implying the presence of a large source of irradiating X-rays impinging on the disc, (iii) evolution of the UVOIR SEDs throughout the mini-outburst, consistent with a truncated irradiated disc cooling and receding further from the BH, and (iv) power-law correlations between optical/UV and X-ray emission, showing only the far-UV emission to be consistent with pure reprocessed X-ray irradiation and evidence for multiple sources of optical and (near and middle) UV emission present. 

We have also presented an investigation of the long-term behaviour of J1753, which, prior to the mini-outbursts, had remained in outburst since its discovery in 2005 \citep[see e.g.][]{Soleri-2013,Shaw-2013a,tomsick2015}. Fitting the X-ray light curve of the initial outburst, we find the classic BH-LMXB outburst and decay profile (i.e. an exponential followed by a linear decay). However, instead of returning to quiescence the source underwent a standstill phase. We attribute this to a variable mass-transfer rate from the companion star \citep[see also][]{coriat2012}, in which, during this phase, $\Dot{M}_2>\Dot{M}_{\rm crit}$ allowing the disc to remain in a hot, ionzied stable state. 

From the long-term X-ray light curve (Fig. \ref{fig:xr_lightcurve}), we calculated the time-averaged $\Dot{M}_2$ to be greater than the critical value during the standstill phase, supporting our hypothesis. The UVOIR SED (obtained during this phase) is well fit with an irradiated disc truncated at large radii, and implies a large fraction of X-rays being reprocessed in the outer disc ($C_{\rm irr}\sim8.8\times10^{-2}$), consistent with the source of irradiating X-rays being produced in a corona existing inside and above the disc. Previous spectral (e.g., \citealt{kajava2016}) and timing (e.g., \citealt{Veledina-2017}) studies during the standstill phase of this source have been found to also support this scenario. Given the available evidence, we suggest that a period of strong X-ray irradiation may have driven the enhanced mass transfer of the standstill phase.

The long-term behaviour observed in J1753 is reminiscent of Z Cam stars, whereby variable mass transfer from the companion drives unusual outbursts, characterized by stalled decays and abrupt changes in luminosity \citep{buatMenard2001,hameury2014}. Thus, we suggest that J1753 is in fact a BH-LMXB analogue to Z Cam type dwarf novae.

\section*{Acknowledgements}
We thank the anonymous referee for useful comments which helped to improve the manuscript. AWS and BET would like to thank P.A. Charles, C.O. Heinke, and G.R. Sivakoff, for useful discussions. AWS and BET are supported by  NSERC Discovery Grants and AWS is also supported by a Discovery Accelerator Supplement. GD acknowledges support from the Centre National d'\'{E}tudes Spatiales (CNES). RMP acknowledges support from Curtin University through the Peter Curran Memorial Fellowship. This research has made use of data, software, and/or web tools obtained from the High Energy Astrophysics Science Archive Research Center (HEASARC), a service of the Astrophysics Science Division at NASA Goddard Space Flight Center (GSFC) and of the Smithsonian Astrophysical Observatory's High Energy Astrophysics Division, and data supplied by the UK Swift Science Data Centre at the University of Leicester. This work has also made extensive use of NASA's Astrophysics Data System (ADS).





\bibliographystyle{mnras}
\bibliography{j1753_refs.bib}



 \appendix
 
 \clearpage
 
 \section{Fits to the UVOIR SEDs}
 
 \begin{figure*}
\subfloat[Epoch 1: February 22-23 (57806-57807)]
  {\includegraphics[width=.46\linewidth,height=.37\linewidth]{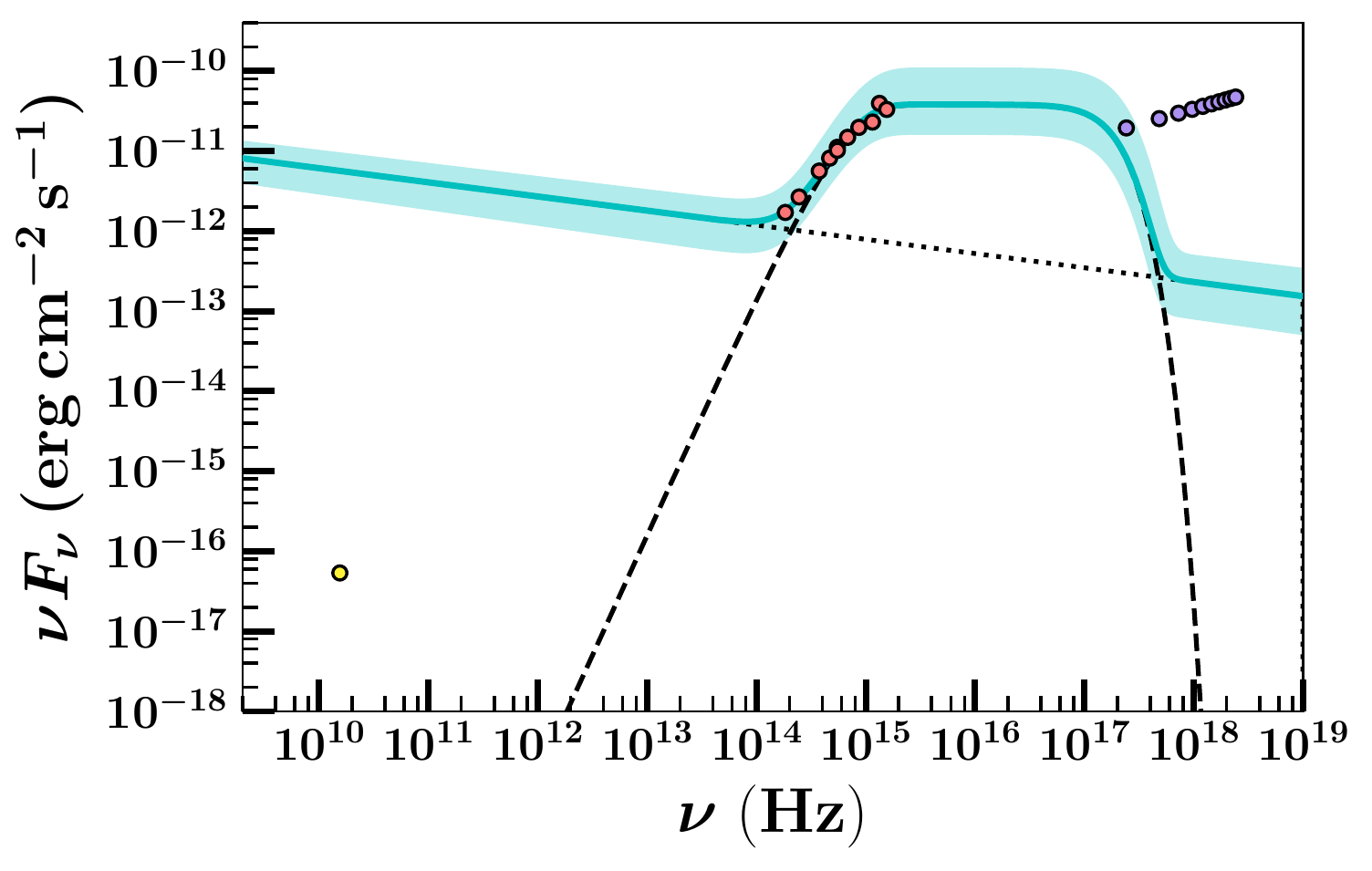}}\hfill
\subfloat[Epoch 2: March 25-26 (57837-57838)]
  {\includegraphics[width=.46\linewidth,height=.37\linewidth]{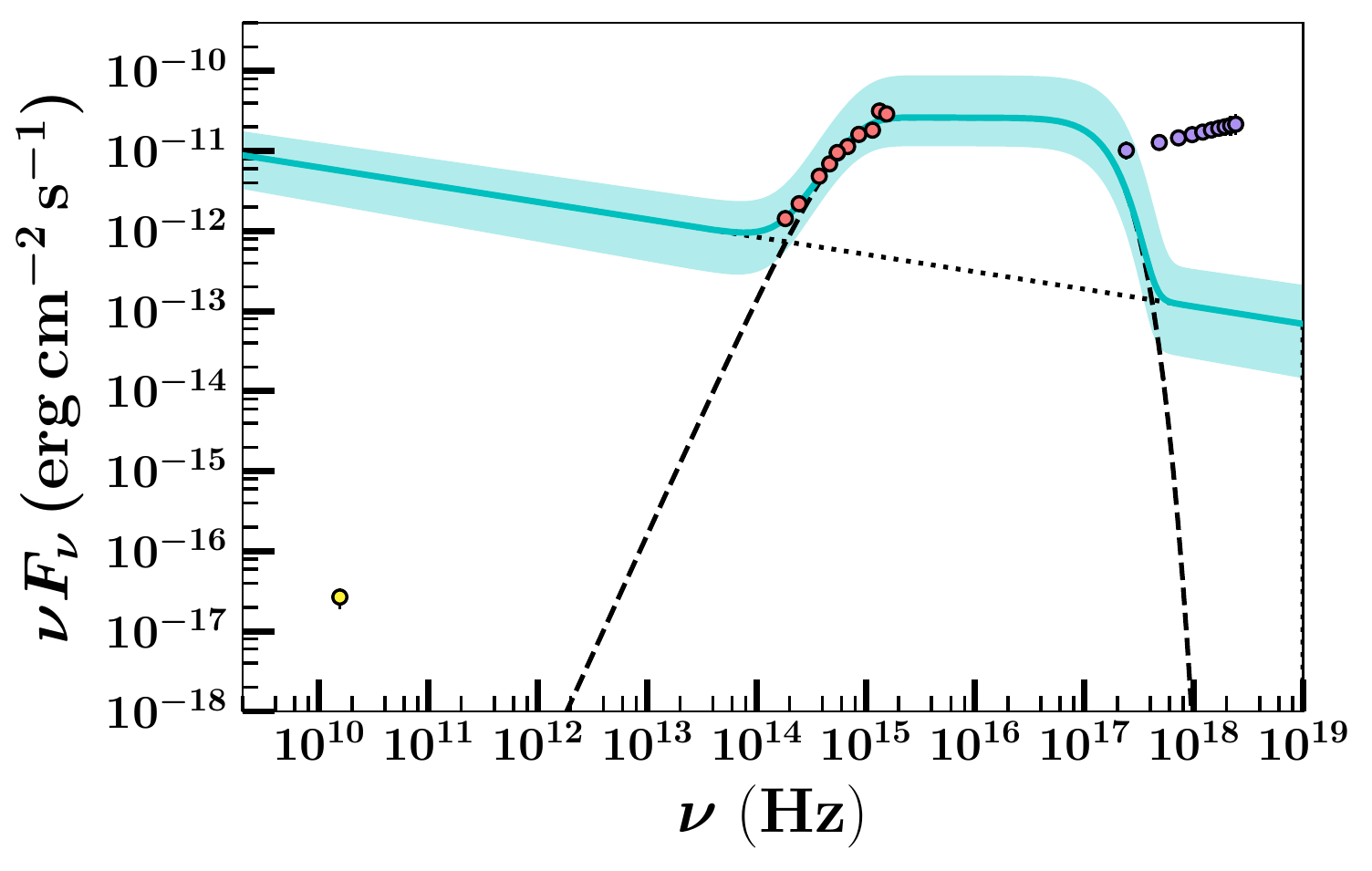}}
  
  \subfloat[Epoch 3: April 1 (57844)]
  {\includegraphics[width=.46\linewidth,height=.37\linewidth]{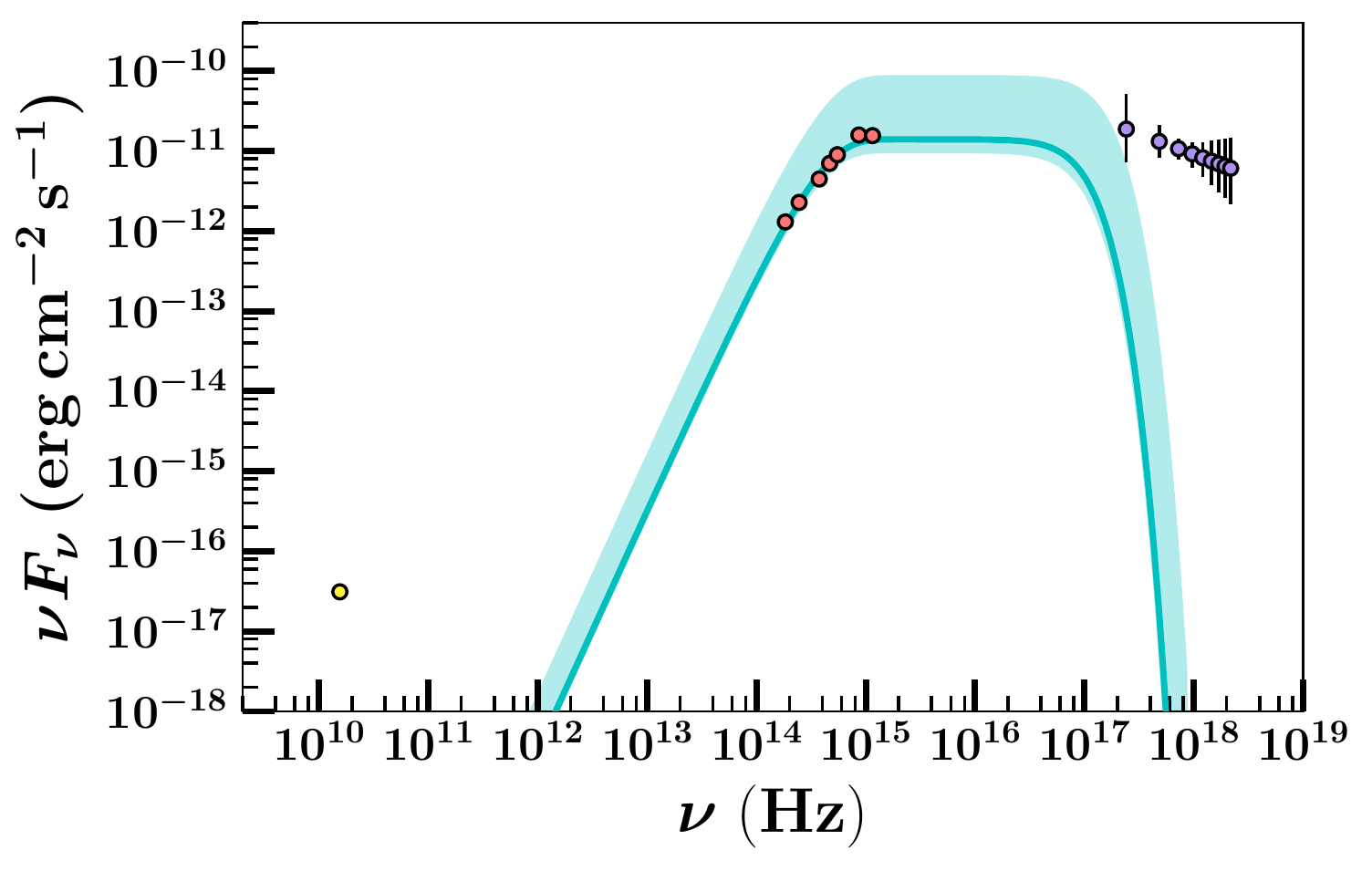}}\hfill
\subfloat[Epoch 4: April 6 (57849)]
  {\includegraphics[width=.46\linewidth,height=.37\linewidth]{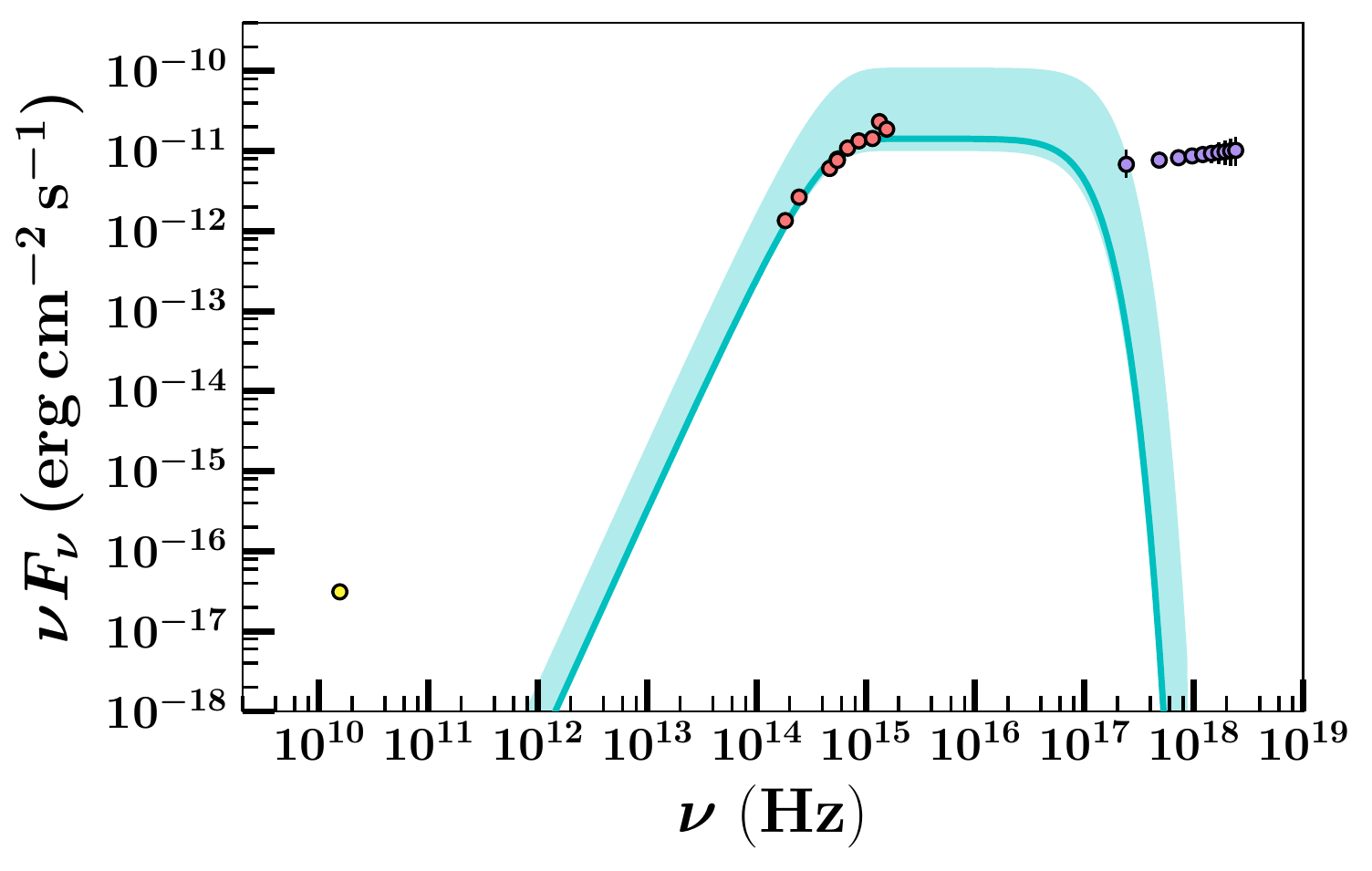}}
  
  \subfloat[Epoch 5: April 8 (57851)]
  {\includegraphics[width=.46\linewidth,height=.37\linewidth]{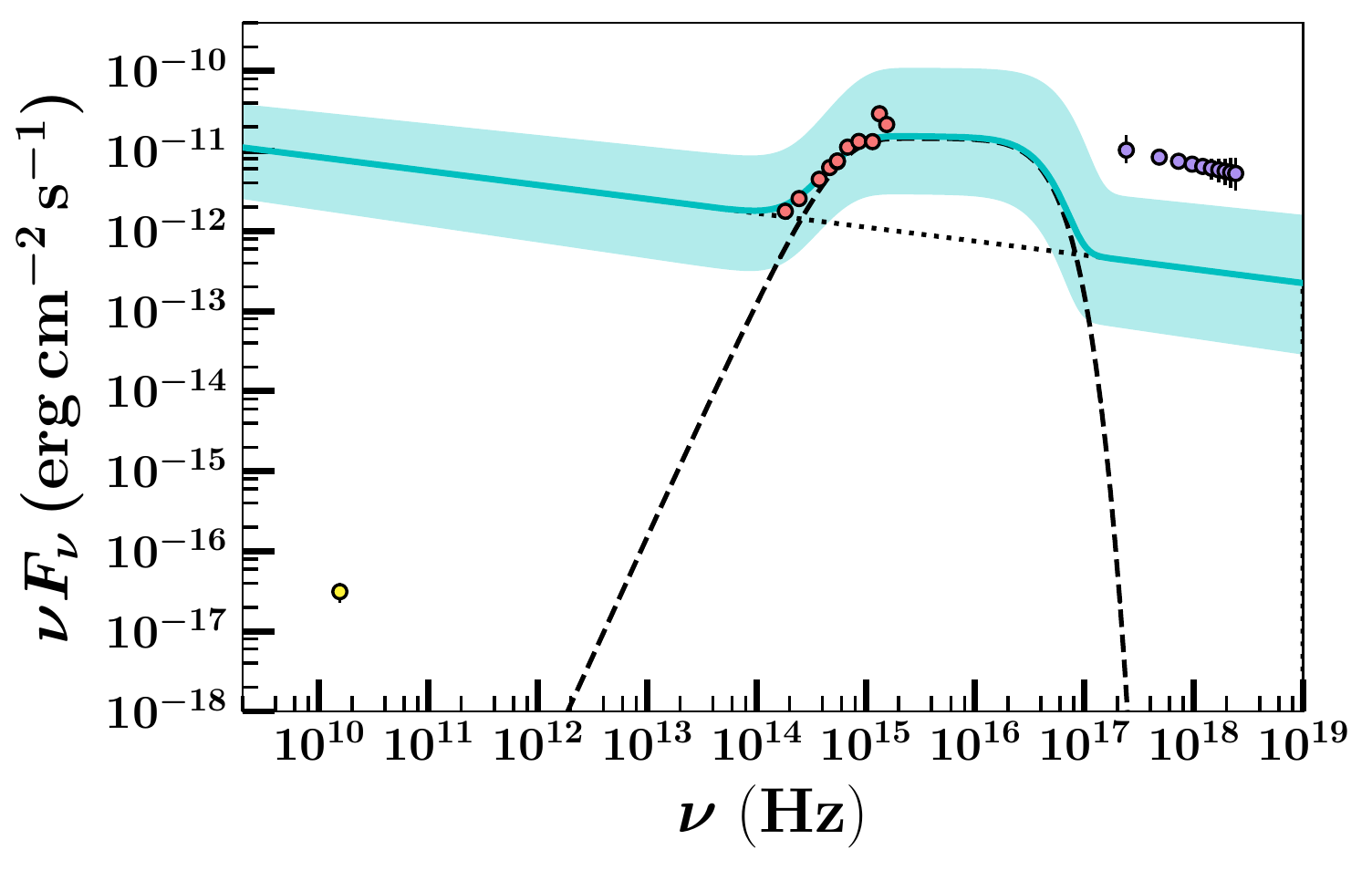}}\hfill
\subfloat[Epoch 6: April 18-19 (57861-57862)]
  {\includegraphics[width=.46\linewidth,height=.37\linewidth]{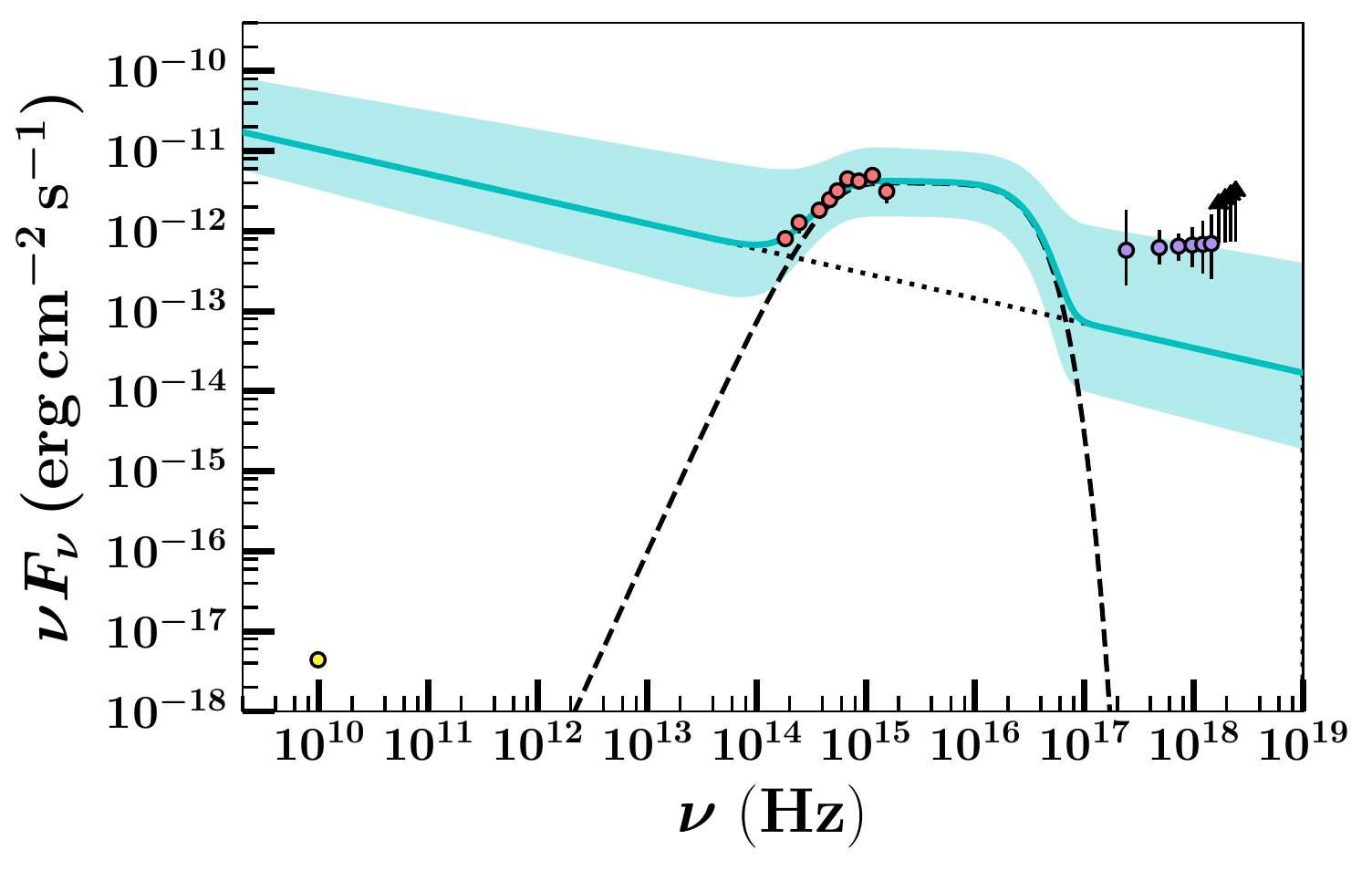}}
  \caption{Broadband SEDs for 6 individual epochs during the Swift\,J1753.5-0127 mini-outburst fit with the irradiated disc model with a temperature distribution $T \propto R^{-1/2}$. All data are simultaneous or quasi-simultaneous (within 1 day) and has been dereddened. The solid cyan line and shaded regions represent the best-fit and 1$\sigma$ confidence intervals from the MCMC fitting algorithm, respectively. The dotted lines show the individual contributions from the irradiated disc (dashed) and jet (dotted), where applicable. Only the UVOIR data ({\em Swift}/UVOT and {\em SMARTS}; red) are fit. X-ray ({\em Swift}/XRT: $2-10$ keV; purple) and radio (VLA: 9.8 GHz and AMI: 15.5 GHz from \citealt{plotkin2017}; yellow) are also plotted to show the multi-wavelength behaviour of the source.}%
  \label{fig:broadband_seds}%
\end{figure*}

\begin{figure*}
\subfloat[Epoch 1: February 22-23 (57806-57807)]
  {\includegraphics[width=.46\linewidth,height=.37\linewidth]{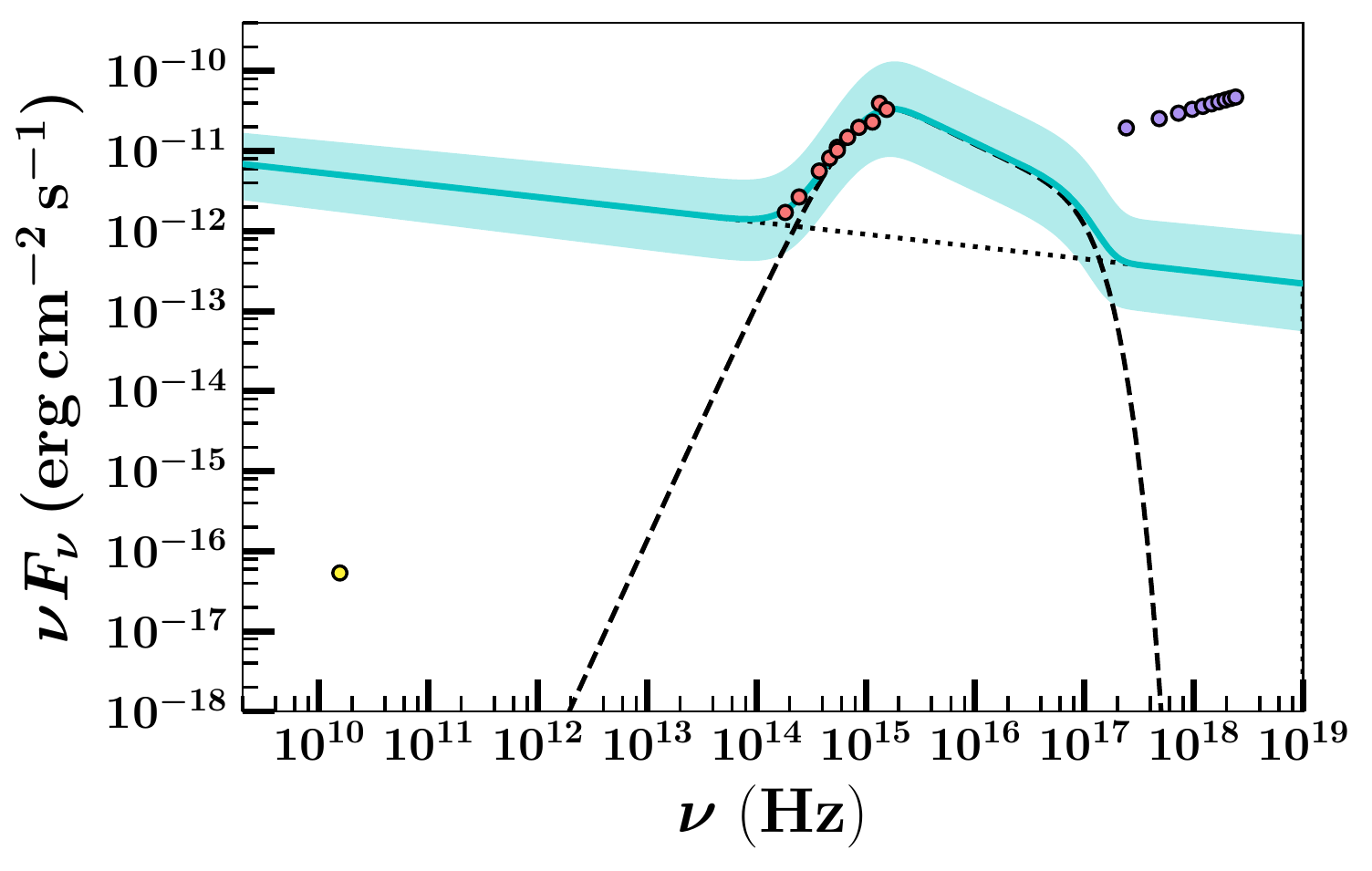}}\hfill
\subfloat[Epoch 2: March 25-26 (57837-57838)]
  {\includegraphics[width=.46\linewidth,height=.37\linewidth]{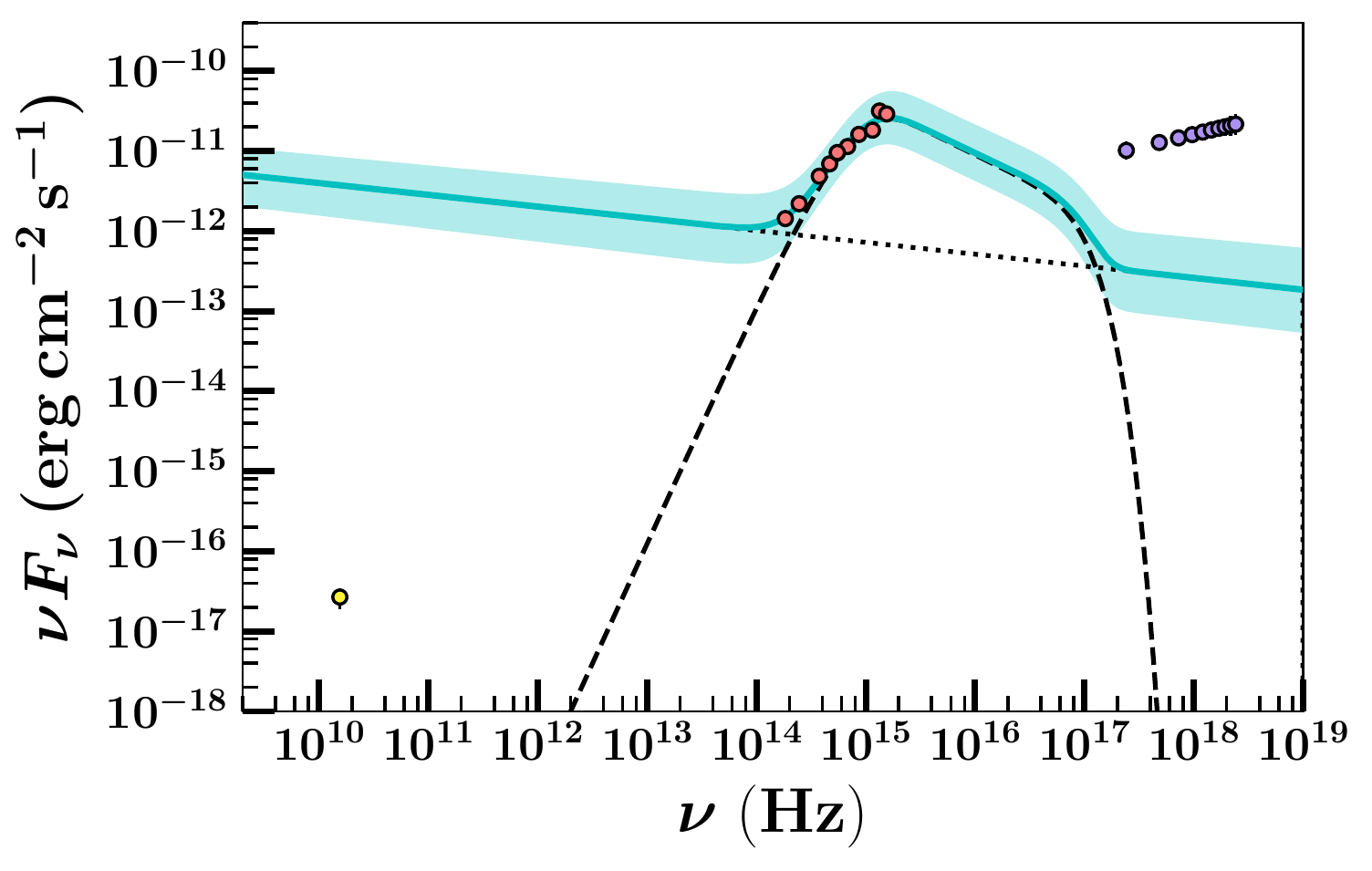}}
  
  \subfloat[Epoch 3: April 1 (57844)]
  {\includegraphics[width=.46\linewidth,height=.37\linewidth]{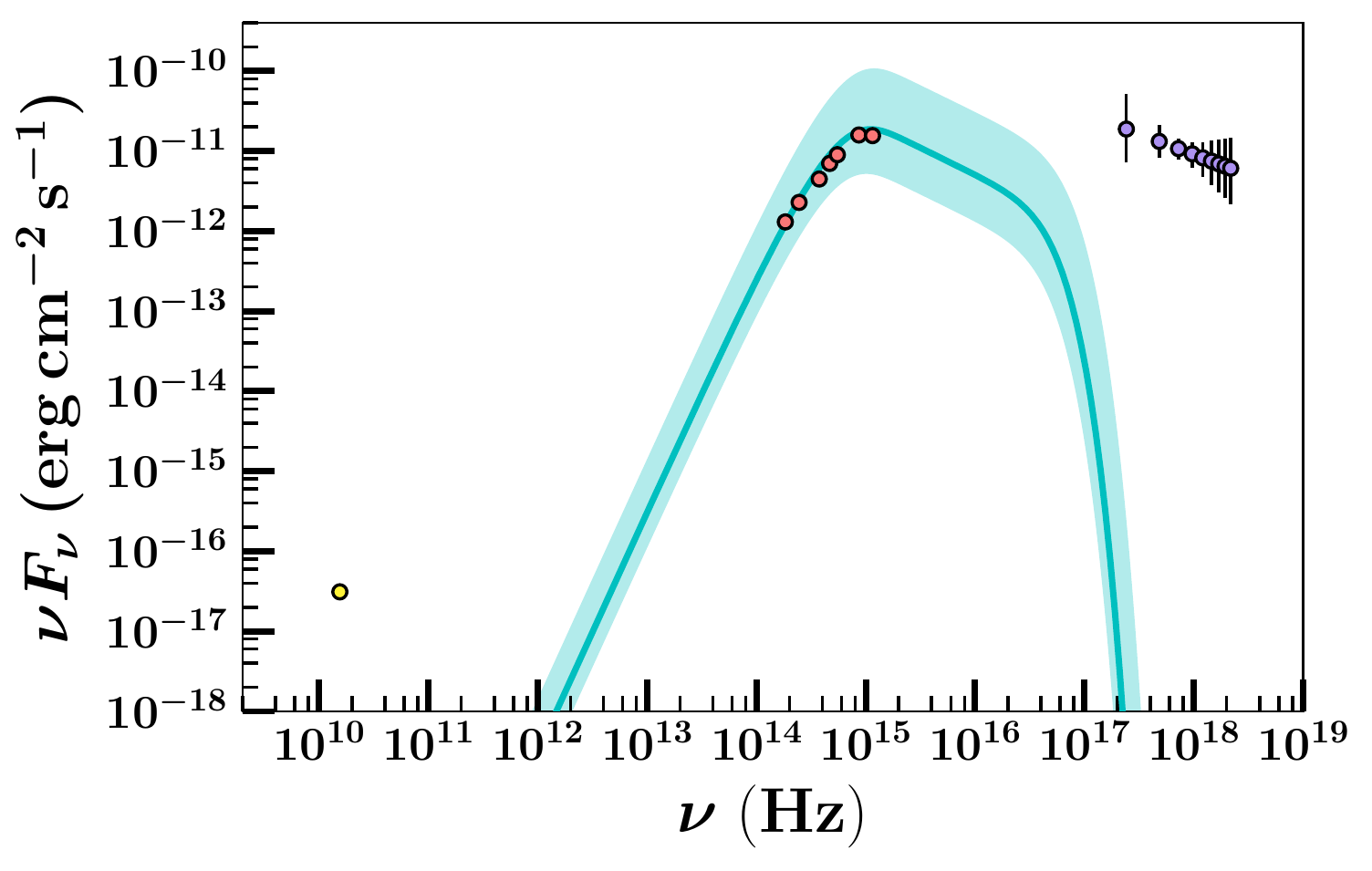}}\hfill
\subfloat[Epoch 4: April 6 (57849)]
  {\includegraphics[width=.46\linewidth,height=.37\linewidth]{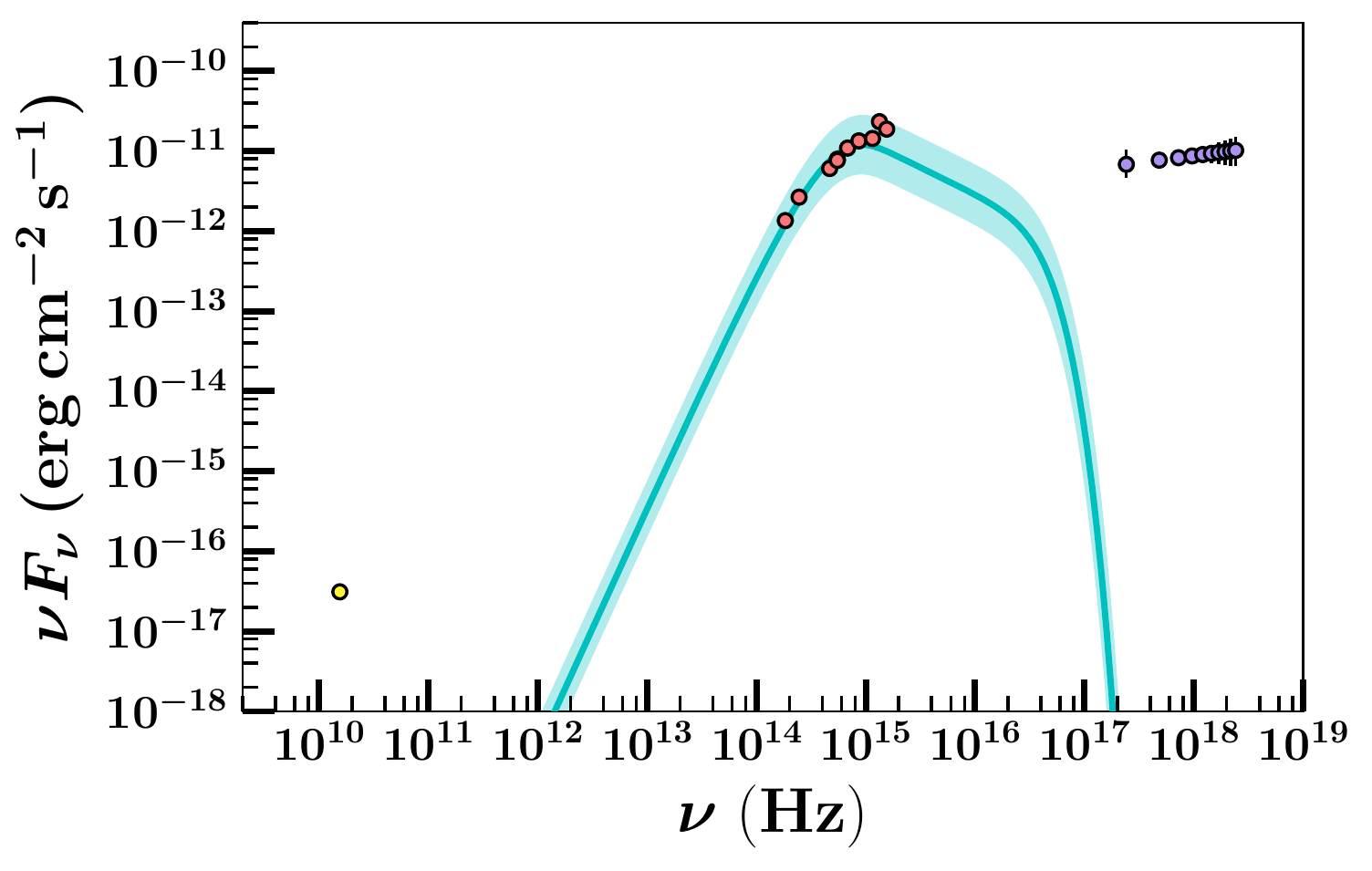}}
  
  \subfloat[Epoch 5: April 8 (57851)]
  {\includegraphics[width=.46\linewidth,height=.37\linewidth]{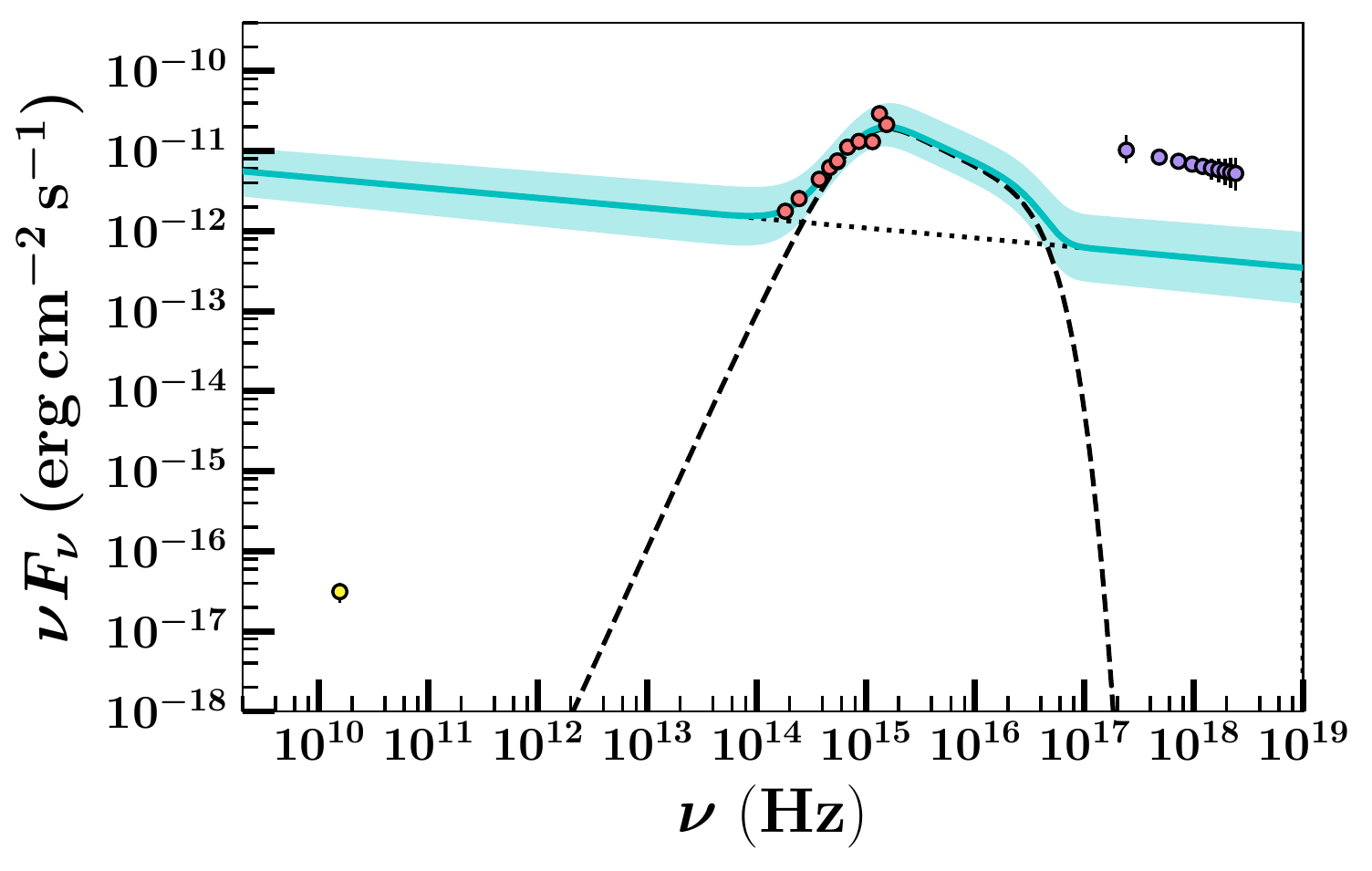}}\hfill
\subfloat[Epoch 6: April 18-19 (57861-57862)]
  {\includegraphics[width=.46\linewidth,height=.37\linewidth]{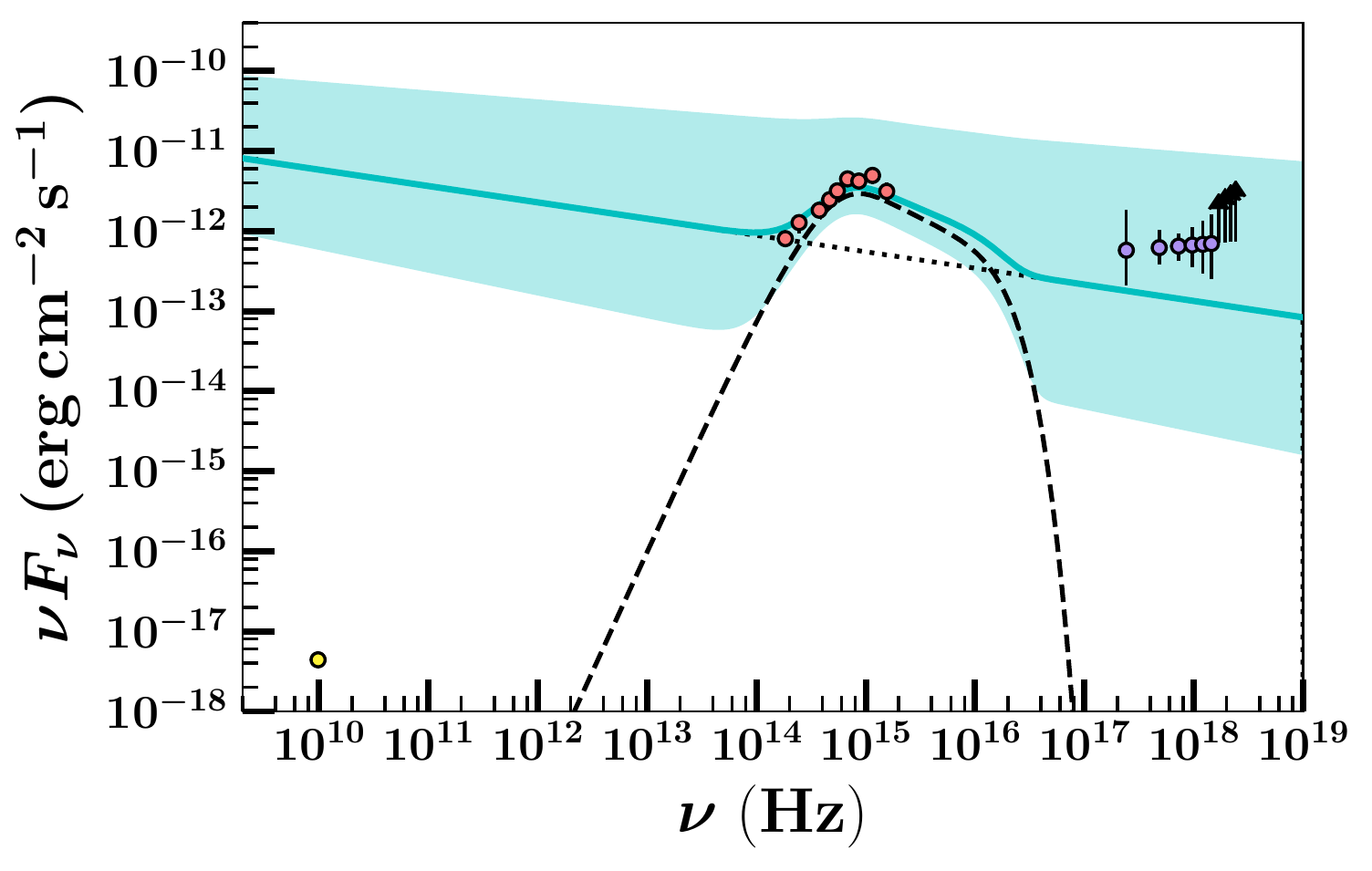}}
  \contcaption{Broadband SEDs for 6 individual epochs during the Swift\,J1753.5-0127 mini-outburst fit with the irradiated disc model with a temperature distribution $T \propto R^{-3/7}$. All data are simultaneous or quasi-simultaneous (within 1 day) and has been dereddened. The solid cyan line and shaded regions represent the best-fit and 1$\sigma$ confidence intervals from the MCMC fitting algorithm, respectively. The dotted lines show the individual contributions from the irradiated disc (dashed) and jet (dotted), where applicable. Only the UVOIR data ({\em Swift}/UVOT and {\em SMARTS}; red) are fit. X-ray ({\em Swift}/XRT: $2-10$ keV; purple) and radio (VLA: 9.8 GHz and AMI: 15.5 GHz from \citealt{plotkin2017}; yellow) are also plotted to show the multi-wavelength behaviour of the source.}%
  \label{fig:broadband_seds2}%
\end{figure*}

\begin{figure*}
\subfloat[Epoch 1: February 22-23 (57806-57807)]
  {\includegraphics[width=.46\linewidth,height=.37\linewidth]{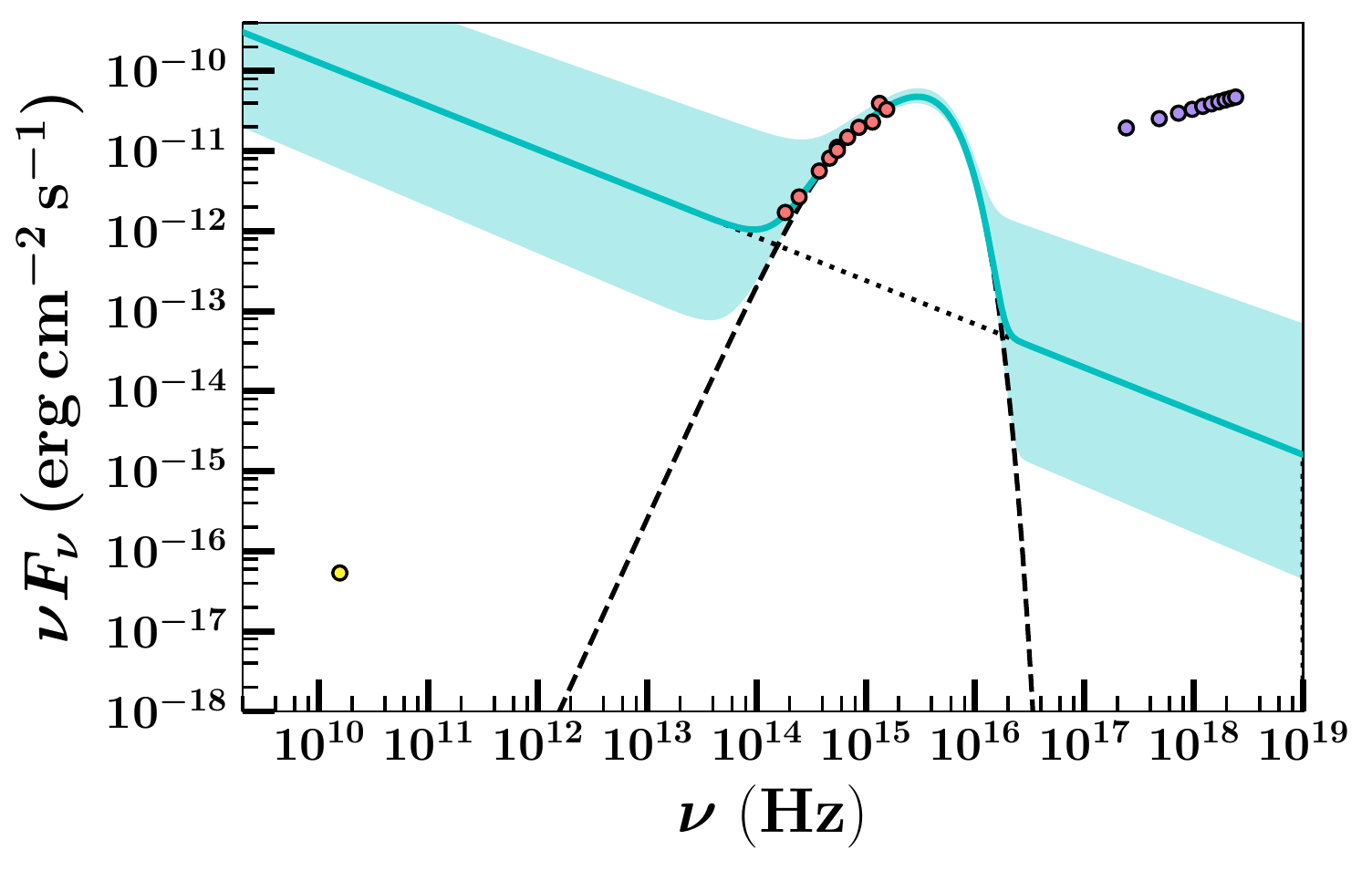}}\hfill
\subfloat[Epoch 2: March 25-26 (57837-57838)]
  {\includegraphics[width=.46\linewidth,height=.37\linewidth]{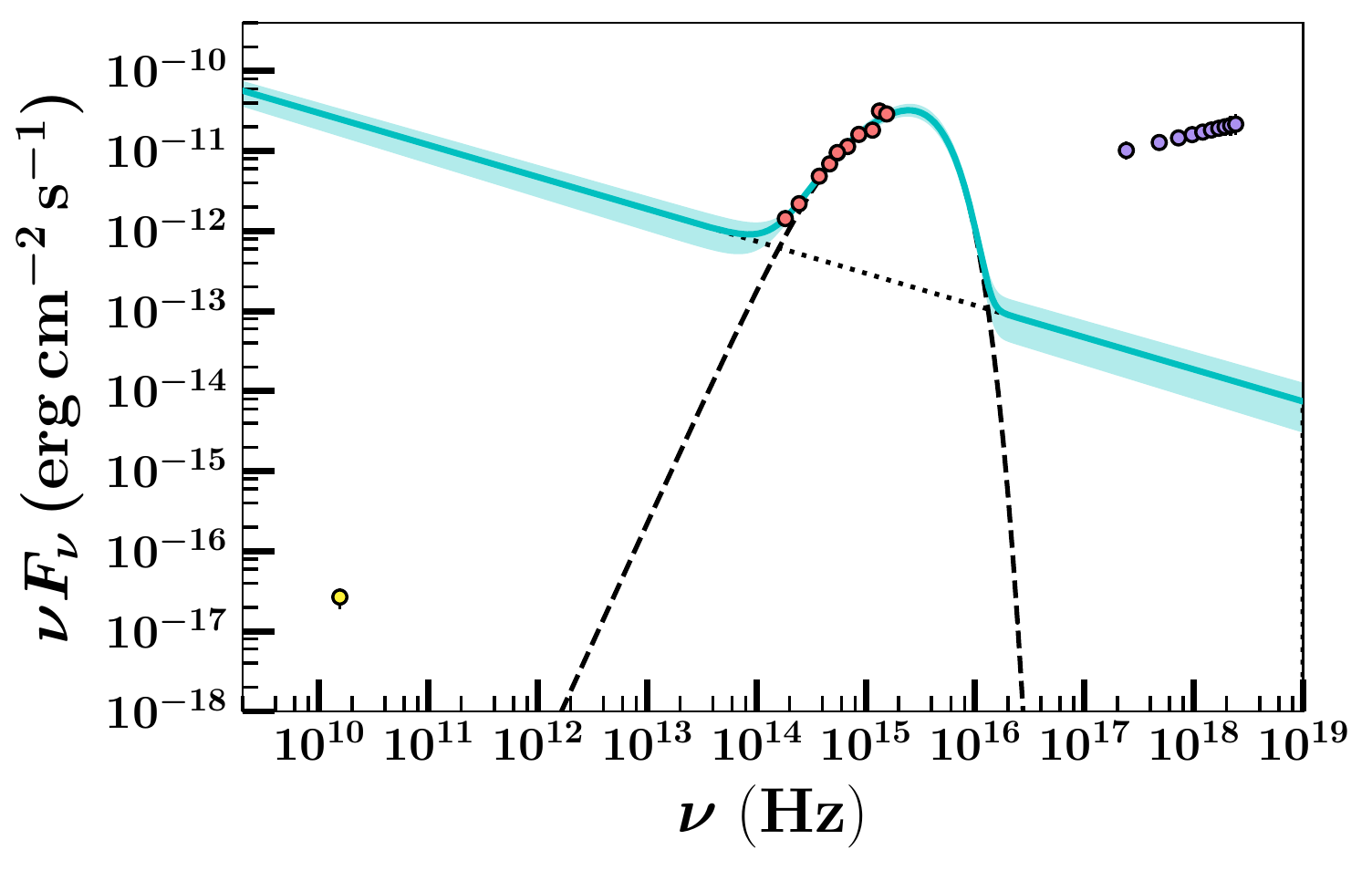}}
  
  \subfloat[Epoch 3: April 1 (57844)]
  {\includegraphics[width=.46\linewidth,height=.37\linewidth]{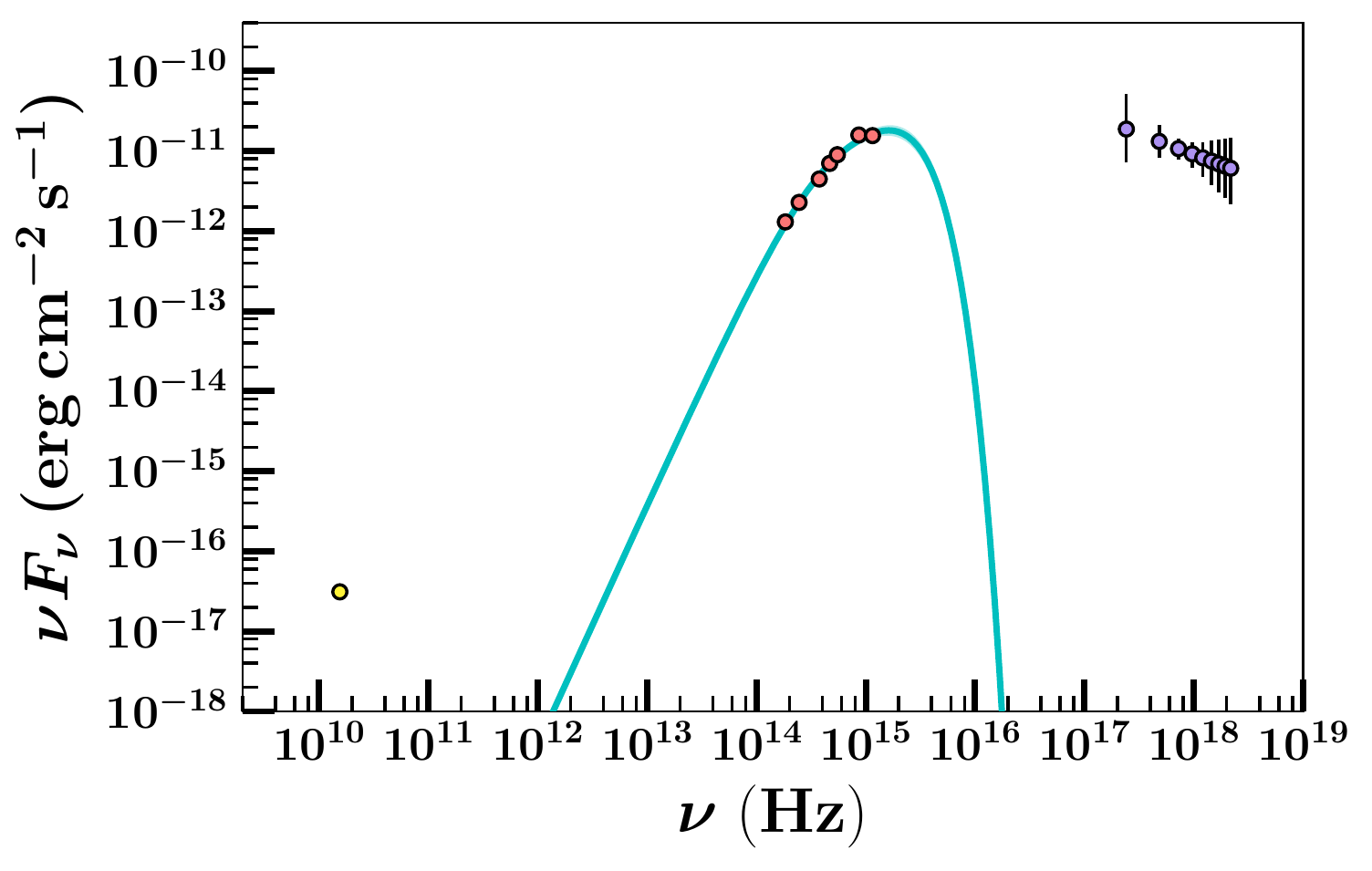}}\hfill
\subfloat[Epoch 4: April 6 (57849)]
  {\includegraphics[width=.46\linewidth,height=.37\linewidth]{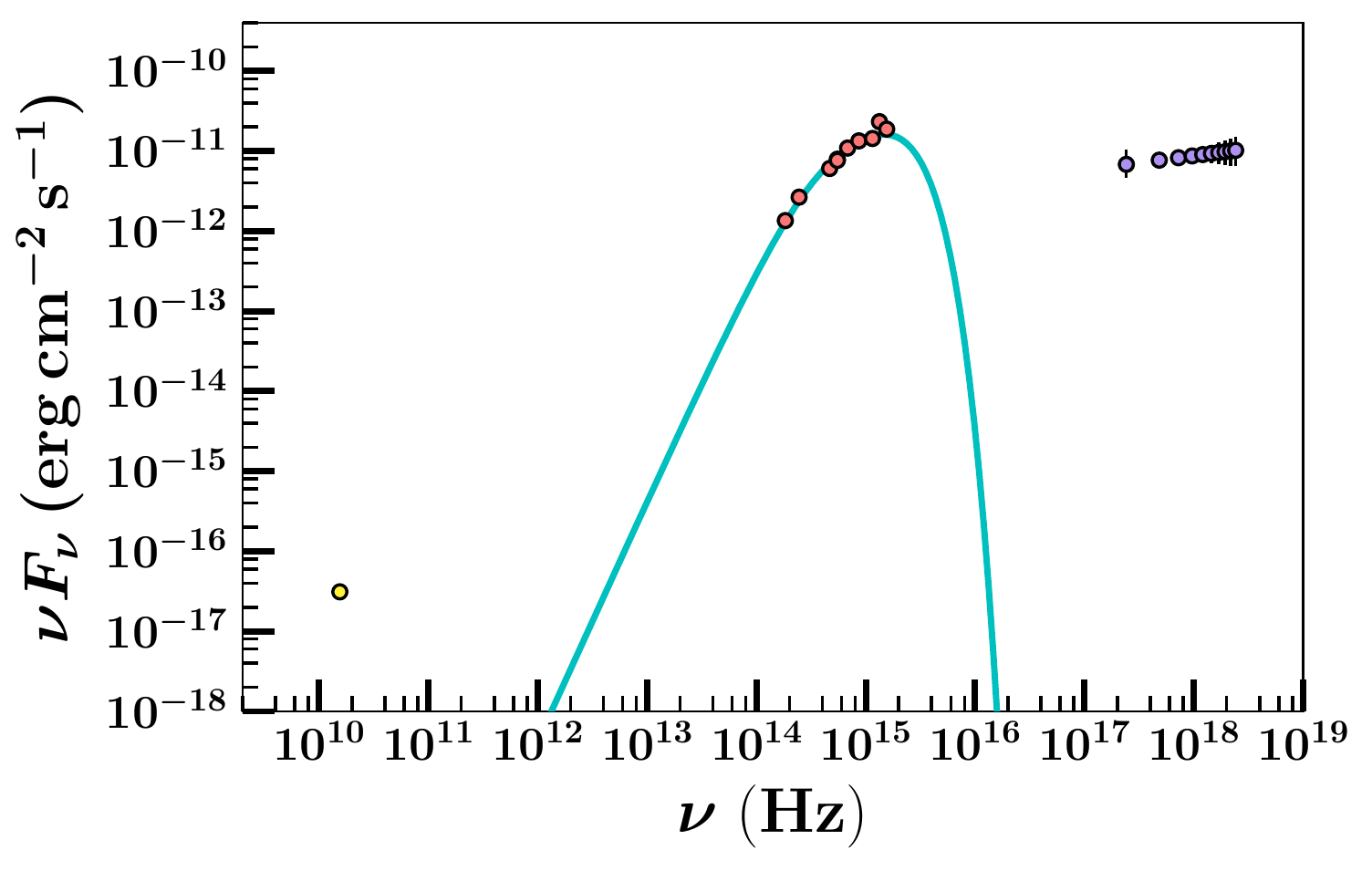}}
  
  \subfloat[Epoch 5: April 8 (57851)]
  {\includegraphics[width=.46\linewidth,height=.37\linewidth]{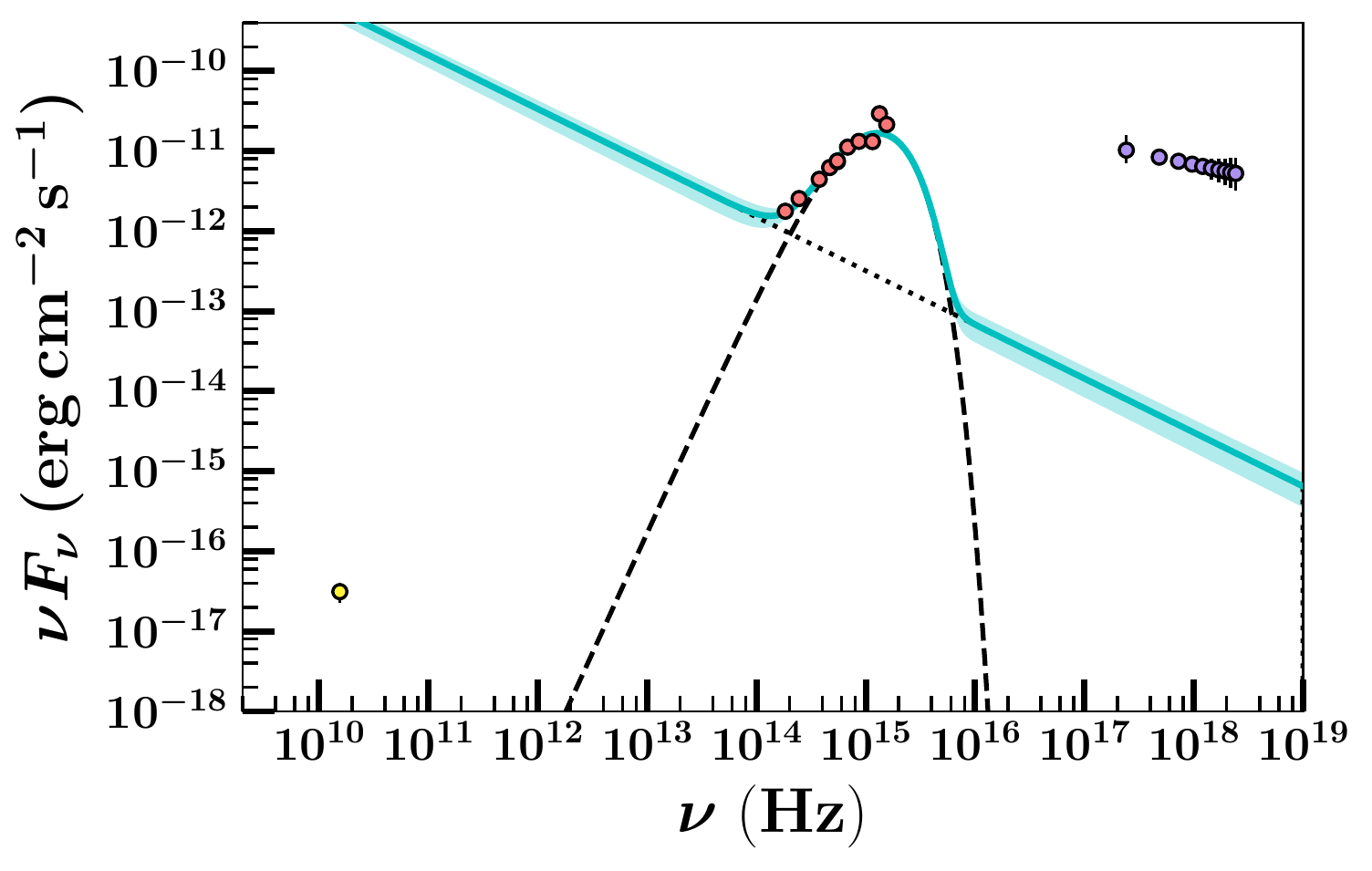}}\hfill
\subfloat[Epoch 6: April 18-19 (57861-57862)]
  {\includegraphics[width=.46\linewidth,height=.37\linewidth]{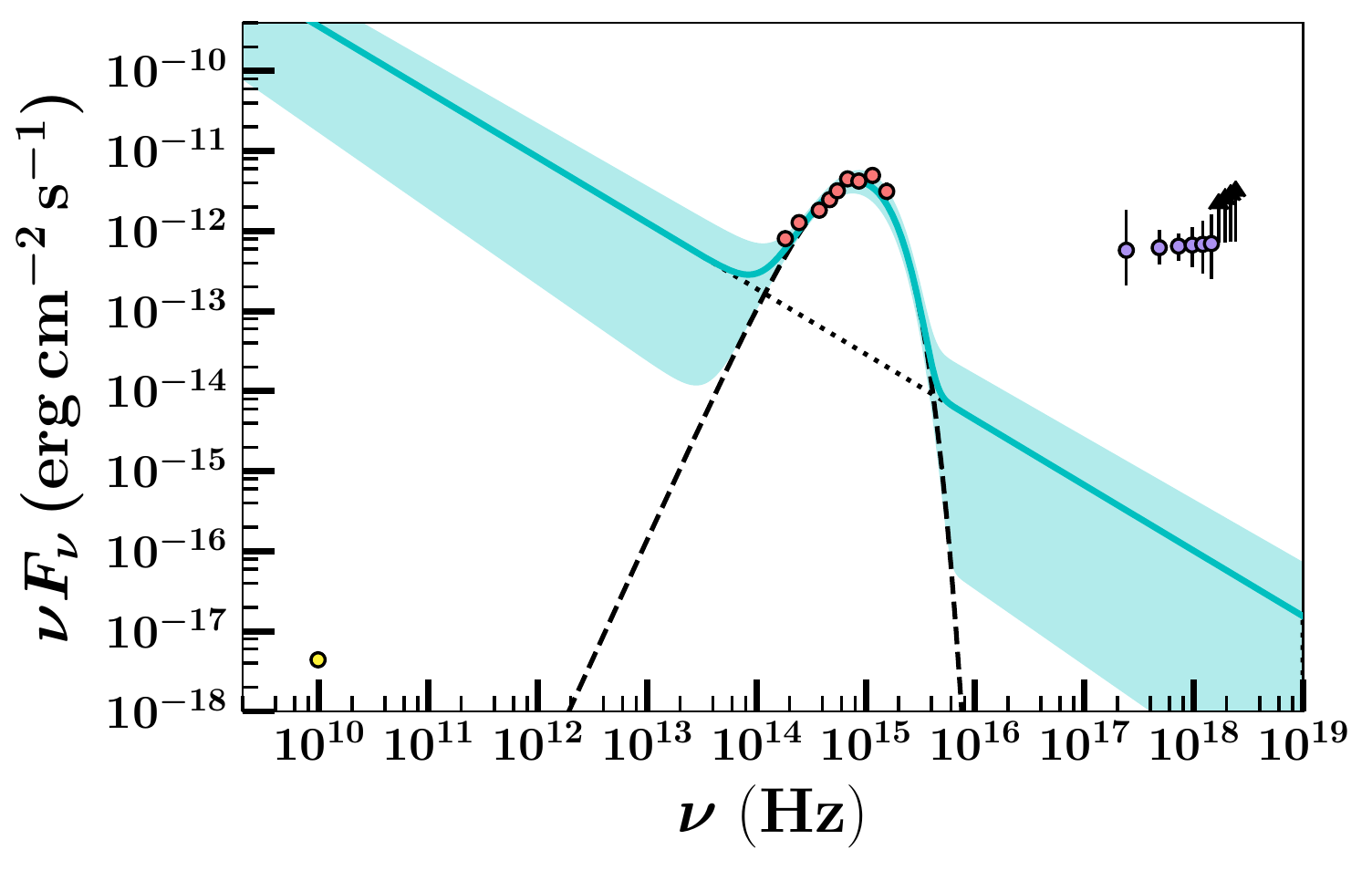}}
  \contcaption{Broadband SEDs for 6 individual epochs during the Swift\,J1753.5-0127 mini-outburst fit with the non-irradiated disc model with a temperature distribution $T \propto R^{-3/4}$. All data are simultaneous or quasi-simultaneous (within 1 day) and has been dereddened. The solid cyan line and shaded regions represent the best-fit and 1$\sigma$ confidence intervals from the MCMC fitting algorithm, respectively. The dotted lines show the individual contributions from the irradiated disc (dashed) and jet (dotted), where applicable. Only the UVOIR data ({\em Swift}/UVOT and {\em SMARTS}; red) are fit. X-ray ({\em Swift}/XRT: $2-10$ keV; purple) and radio (VLA: 9.8 GHz and AMI: 15.5 GHz from \citealt{plotkin2017}; yellow) are also plotted to show the multi-wavelength behaviour of the source.}%
  \label{fig:broadband_seds3}%
\end{figure*}

\clearpage


  
  
  



\bsp	
\label{lastpage}
\end{document}